\documentclass[sigconf]{acmart}
\AtBeginDocument{%
  }

\copyrightyear{2026}
\acmYear{2026}
\setcopyright{cc}
\setcctype{by}
\acmConference[CHI '26]{Proceedings of the 2026 CHI Conference on Human Factors in Computing Systems}{April 13--17, 2026}{Barcelona, Spain}
\acmBooktitle{Proceedings of the 2026 CHI Conference on Human Factors in Computing Systems (CHI '26), April 13--17, 2026, Barcelona, Spain}
\acmPrice{}
\acmDOI{10.1145/3772318.3790303}
\acmISBN{979-8-4007-2278-3/2026/04}


\usepackage{tabularx}

\usepackage{subcaption}
\usepackage{multirow}

\def\anon{0}
\newcommand{\anonymize}[1]{
    \ifthenelse{\anon > 0}{[redacted for blind review]}{#1}
}
\newcommand{\revision}[1]{#1} 

\begin{document}

\title{When Workout Buddies Are Virtual: AI Agents and Human Peers in a Longitudinal Physical Activity Study}

\author{Alessandro Silacci}
\orcid{0000-0001-8121-3013}
\affiliation{%
\department{Persuasive Technology Lab}
\institution{University of Lausanne}
\city{Lausanne}
\country{Switzerland}}
\affiliation{%
\institution{School of Management of Fribourg, HES-SO University of Applied Sciences and Arts Western Switzerland}
\city{Fribourg}
\country{Switzerland}}
\email{alessandro.silacci@unil.ch}

\author{Mauro Cherubini}
\orcid{0000-0002-1860-6110}
\affiliation{%
\department{Persuasive Technology Lab}
\institution{University of Lausanne}
\city{Lausanne}
\country{Switzerland}}
\email{mauro.cherubini@unil.ch}
  
\author{Arianna Boldi}
\orcid{0000-0002-2825-3443}
\affiliation{%
 \department{Department of Psychology}
 \institution{University of Turin}
 \city{Torino}
 \country{Italy}}
\email{arianna.boldi@unito.it}

\author{Amon Rapp}
\orcid{0000-0003-3855-9961}

\affiliation{%
 \department{Computer Science Department}
  \institution{University of Turin}
  \city{Torino}
  \country{Italy}}
\email{amon.rapp@unito.it}

\author{Maurizio Caon}
\orcid{0000-0003-4050-4214}
\affiliation{%
\institution{School of Management of Fribourg, HES-SO University of Applied Sciences and Arts Western Switzerland}
\city{Fribourg}
\country{Switzerland}}
\email{maurizio.caon@hes-so.ch}

\renewcommand{\shortauthors}{Silacci et al.}

\begin{abstract}
Physical inactivity remains a critical global health issue, yet scalable strategies for sustained motivation are scarce. Conversational agents designed as simulated exercising peers (SEPs) represent a promising alternative, but their long-term impact is unclear. We report a six-month randomized controlled trial (N=280) comparing individuals exercising alone, with a human peer, or with a large language model-driven SEP. Results revealed a partnership paradox: human peers evoked stronger social presence, while AI peers provided steadier encouragement and more reliable working alliances. Humans motivated through authentic comparison and accountability, whereas AI peers fostered consistent, low-stakes support. These complementary strengths suggest that AI agents should not mimic human authenticity but augment it with reliability. Our findings advance human-agent interaction research and point to hybrid designs where human presence and AI consistency jointly sustain physical activity.
\end{abstract}


\maketitle

\section{Introduction}
\label{sec:introduction}
Physical inactivity remains a pressing global health challenge, and scalable approaches to sustained motivation are urgently needed \cite{world_health_organization_global_2009, world_health_organization_global_2022}. Decades of behavioral science show that social support is one of the strongest predictors of physical activity \cite{ryan_self-determination_2018, deci_what_2000, bandura_social_1986}. \revision{A central mechanism behind the effectiveness of social support is relatedness -- the sense of feeling connected, supported, and accompanied by another person. Within Self-Determination Theory, relatedness is a core psychological need that helps sustain motivation for physical activity \cite{ryan_self-determination_2018,ryan_intrinsic_2020,deci_what_2000,molina_motivation_2023}. When people feel that someone is ``with them,'' they are more willing to persist, challenge themselves, and recover from setbacks \cite{avraham_determinants_2024,bouten_students_2025,bong_good_2023}.} Support from peers in particular can spark motivation through comparison, accountability, and encouragement \cite{laird_role_2016, duncan_sources_2005, cao_associations_2023, van_luchene_influence_2021, lindsay_smith_association_2017, sarkar_social_2016}. Yet these same dynamics are fragile: human peers may be unavailable, disengage, or inadvertently demotivate \cite{molina_motivation_2023}. The challenge is especially acute for young adults, who navigate shifting social circles and variable sources of support \cite{cao_associations_2023, van_luchene_influence_2021}.

\par To fill these gaps, researchers have increasingly explored conversational agents -- such as chatbots and embodied agents -- as scalable ways to provide data-driven social support \cite{luo_promoting_2021, bickmore_automated_2013, kocielnik_reflection_2018, de_freitas_ai_2024}. These agents can offer encouragement, feedback, and adaptive goals around the clock \cite{zhang_artificial_2020, de_freitas_ai_2024}. Advances in Large Language Models (LLMs) now enable more fluid and persuasive interactions \cite{bai_llm-generated_2025, schoenegger_large_2025, li_systematic_2023}, raising expectations that AI peers might sustain motivation in ways that earlier systems could not. Yet questions remain about their long-term effectiveness and the nature of the human-agent relationships they can foster \cite{luo_promoting_2021}. \revision{A key determinant to how users interpret and relate to virtual agents is visual embodiment: appearance can signal humanness, artificiality, warmth, or distance, which in turn shapes believability and the expectations users bring into the interaction \cite{bogdanovych_formalising_2015, niewiadomski_warmth_2010, ijaz_enhancing_2011}. How different visual designs influence users’ sense of connection and support therefore remains an important open question.}

\par One promising direction is to reconceptualize agents not as coaches, but as \textit{Simulated Exercising Peers (SEPs)} -- co-participants who appear to exert effort alongside the user \cite{silacci_navigating_2025, silacci_when_2025}. This design embraces the social mechanisms of comparison and companionship rather than authority. However, it also raises new uncertainties: can an artificial peer be perceived as authentic? Will users accept its encouragement as meaningful support? And can SEPs sustain motivation over the long term, beyond the initial novelty effect?

\par \revision{To address these uncertainties and clarify the relational potential of SEPs, we formulated the following research questions:}
\begin{itemize}
    \item \textbf{RQ1:} To what extent can AI agents effectively provide relatedness support to promote and sustain physical activity?
    \item \textbf{RQ2:} How do AI agents compare to humans in providing relatedness support to promote and sustain physical activity?
    \item \textbf{RQ3:} How does the visual appearance of an AI agent, specifically a human figure versus a cyborg, affect its ability to provide relatedness support?
    \item \textbf{RQ4:} How do users actively assess and evaluate the virtual peers' visual appearance, behavior, and communication style, considering both their level of believability and the potential for perceived deception within the context of a physical activity intervention?
\end{itemize}

\par \revision{To examine these questions, we carried out a six-month randomized controlled trial (N = 280) with four conditions: human–human dyads (HUM), AI peers with either a human-like avatar (SEPH) or a cyborg-like avatar (SEPC), and a no-peer control (CON). Our study makes three contributions. First, we provide one of the first long-term empirical evaluations of an LLM-powered SEP in supporting physical activity. Second, we examine how human and AI peers differ in fostering motivation and relational dynamics. In doing so, we uncover a key tension we term the \textit{partnership paradox}: while human peers foster a stronger sense of social presence, AI peers prove more reliable at establishing a task-focused working alliance. Third, through a mixed-methods analysis, we show how an agent’s appearance and conversational style shape perceptions of believability and authenticity, offering design insights for more effective and engaging AI-driven health interventions.}

\par Before delving into the study design, we are going to review the related literature which supported the proposed research questions and the hypotheses tested in the presented study.
\section{Related work}
\label{sec:related_work}

\subsection{Social support and physical activity}
\label{sub:rw_social_support}
Social support is a cornerstone of behavior change interventions for physical activity (PA) \cite{orji_persuasive_2018, almutari_how_2019}. Research rooted in Self-Determination Theory (SDT) \cite{ryan_self-determination_2018} and Social Cognitive Theory (SCT) \cite{bandura_social_1986} demonstrates that social connections or social environments play a key role in the shaping of new motivation or behavior. Social connectedness, from an SDT point of view, enhances intrinsic motivation by fulfilling one's basic psychological need for relatedness \cite{ryan_self-determination_2018, ryan_intrinsic_2020}. Within this framework, peers play a particularly critical role. Unlike authority figures (e.g., coaches, teachers), peers motivate through horizontal relationships that foster collaboration, competition, and mutual accountability \cite{avraham_determinants_2024, bouten_students_2025}. Social comparison theory further explains these effects: individuals are most motivated when comparing themselves to ``similar others'' \cite{festinger_theory_1954, barta_similar_2023}. However, this dynamic can be double-edged -- competition and comparison can spark motivation but also risk discouragement if not managed carefully \cite{molina_motivation_2023}. 

\par Prior work in education and exercise contexts highlights the divergent motivational impacts of vertical versus horizontal support. Authority-driven support tends to foster compliance, while peer-driven support is better at sustaining engagement through shared experience \cite{bouten_students_2025, bong_good_2023}. For young adults in particular, peer influence is a strong determinant of forming and maintaining PA habits \cite{avraham_determinants_2024}. Research shows that positive social connections are a primary motivator for sustained PA, and that support from family and friends enhances self-efficacy and engagement \cite{deng_influence_2023, scarapicchia_social_2017}. While human peers are powerful motivators, their availability and reliability are inconsistent.


\par At the same time, interactive features in fitness applications have been shown to foster relatedness between users \cite{molina_motivation_2023,ren_exploring_2018}, and this relatedness in turn strengthens intrinsic motivation. Even when mediated through distant, online connections, peers can influence physical activity by sharing goals, comparing progress, or encouraging each other with positive messages \cite{ren_exploring_2018, ebrahimi_characterizing_2016,sheridan_systematic_2014}. \revision{Empirical work shows that social interaction during exercise strengthens relatedness need satisfaction and supports sustained PA \cite{zhang_effect_2022}. HCI studies similarly highlight that digital tools shape users’ sense of social connection \cite{wong_its_2025}, and critiques of gamified PA systems caution that extrinsic rewards rarely sustain motivation without supporting basic psychological needs \cite{postma_cost_2023, cherubini_unexpected_2020}. Recent findings further show that people increasingly treat AI health tools as collaborative partners \cite{van_arum_selective_2025}, suggesting that AI agents in PA platforms may help fulfill users’ relatedness needs -- a possibility that warrants additional empirical studies.}


\subsection{AI agents in health behavior change}
\label{sub:rw_va_health_behavior_change}

To provide scalable, consistent support, researchers have long turned to virtual agents. The predominant paradigm is the agent-as-coach, where an AI agent guides users through feedback, assessment, and goal-setting \cite{beinema_tailoring_2021, kamali_virtual_2020, watson_internet-based_2012}. This design aligns with traditional vertical support models, and relational agents in this role have shown promise in building trust and engagement \cite{bickmore_establishing_2005, lisetti_i_2013}. Yet the coach role emphasizes expertise rather than mutual participation, raising the question of whether agents can also act as peers.

\par The agent-as-peer paradigm is comparatively underexplored but increasingly relevant in HCI. Whereas the coach role emphasizes expertise and authority, the peer role highlights similarity, co-experience, and mutual accountability -- collaborating, competing, and encouraging through shared effort. In contexts where human peers are unavailable, agents can provide a mediated sense of social presence comparable to that offered by peers \cite{griffiths_exercise_2018}. Designed as companions, such agents can nurture social-emotional relationships, increase relational depth, and support longer-term bonds \cite{bickmore_establishing_2005,bickmore_its_2005}. In doing so, they foster engagement, trust, and a sense of connection similar to that experienced with friends or partners \cite{bickmore_its_2005,sillice_using_2018}, reducing feelings of loneliness \cite{de_freitas_ai_2024}. Their non-judgmental nature fosters users' self-disclosure, providing psychological support while also promoting adherence and motivation \cite{bickmore_establishing_2005,bickmore_its_2005,sillice_using_2018}. Similar effects have been observed in empathic embodied conversational agents, where adaptive verbal and nonverbal behaviors increase acceptance and engagement in behavior change interventions \cite{lisetti_i_2013}. Importantly, these relational qualities may help mitigate the declines in intervention effectiveness often observed once novelty fades \cite{oh_enhancing_2025}. Yet significant barriers remain. Authenticity is undermined when agents are explicitly identified as AI, reducing their legitimacy as peers \cite{yin_ai_2024}. Believability hinges on perceived effort, with users dismissing comparisons that do not reflect genuine exertion \cite{silacci_when_2025}. Acceptance is also domain-dependent: while physical activity contexts appear receptive to AI peers, other health domains may not. Together, these insights highlight both the promise and fragility of agent-as-peer designs.

\par Recent advances in Large Language Models (LLMs) amplify both the potential and challenges of agent design. LLMs enable fluid, personalized dialogue and persuasive messaging \cite{schoenegger_large_2025, bai_llm-generated_2025}. Yet recent work on AI risk highlights how LLM-based approaches can produce vague, incomplete, or misleading accounts that undermine trust and transparency \cite{rao_riskrag_2025}. To date, most evaluations remain short-term and feasibility-focused, leaving open the critical question of whether LLM-powered peers can sustain motivation over time. 

\par Together, these findings highlight both the potential and the challenges of designing AI peers. These challenges are best understood through the lens of human-agent interaction, which we turn to next.


\subsection{Human-agent interaction: believability, expectations, and social presence}
\label{sub:rw_key_concepts_hai}

Believability, rather than realism, is central to human-agent interaction. A believable agent is not one that looks perfectly human, but one whose behavior is coherent and life-like in context \cite{bogdanovych_formalising_2015}. Such believability stems from qualities like personality, emotion, self-motivation, and contextual awareness, which enable users to suspend disbelief even when they know the agent is artificial \cite{bogdanovych_formalising_2015}. Social-emotional cues also matter: agents that convey warmth and competence are perceived as more believable, highlighting the importance of relational affect over surface realism \cite{niewiadomski_warmth_2010}. This aligns with the notion of awareness believability, where agents that appear sensitive to their environment, internal state, and social context are judged as more credible partners \cite{ijaz_enhancing_2011}. In the context of physical activity, believability further depends on aligning an agent’s aesthetics, behavior, and communication style with user expectations, as mismatches can produce deception, disengagement, or uncanny feelings \cite{silacci_navigating_2025}. \revision{Such incoherence disrupts how users interpret the agent’s role and undermines its capacity to provide meaningful support -- making expectation alignment central to both relatedness and believability \cite{silacci_navigating_2025}.} More recently, generative agents demonstrate how memory, reflection, and planning architectures allow LLM-powered systems to sustain socially coherent behavior that users experience as believable \cite{park_generative_2023}. Together, these perspectives suggest that believability is less about mimicking human form and more about behavioral consistency, contextual responsiveness, and social authenticity.

\par While believability emphasizes coherence, humanness highlights the social and expressive cues that shape how users relate to an agent. Subtle demonstrations of human‑like features \revision{-- such as naturalistic response delays with ``typing'' indicators or brief social cues like self-reference or polite acknowledgements --} can increase satisfaction in human-agent interaction \cite{brendel_paradoxical_2023}. Such mimicry strengthens perceptions of agents as social actors \cite{brendel_paradoxical_2023} and encourages anthropomorphization, a natural human tendency when engaging with non‑human entities \cite{rapp_how_2025,nass_machines_2000}. Higher levels of humanness are associated with increased trust, closeness, engagement, and social presence \cite{rapp_how_2024,rapp_human_2021,chaves_how_2021}. They can also reduce user aggression borne of frustration, as more natural and fluid conversations shift interaction away from rigid, command‑driven exchanges \cite{brendel_paradoxical_2023,cowan_what_2017,rapp_how_2025}. However, pushing too far toward human‑likeness risks triggering uncanny valley effects, undermining acceptance and trust \cite{mori_uncanny_1970,rapp_human_2021,he_plan_2025}. This risk is exacerbated in LLM‑based systems, which may produce nonsensical or unexpected responses (i.e., hallucinations) that unsettle users and erode believability \cite{rapp_how_2025}. This concern directly connects to expectation management, where the degree of humanness shapes what users anticipate from the interaction.

\par Social presence is the experiential outcome of believability and humanness, capturing the extent to which users feel they are ``with'' a social partner. Classic work shows that even minimal cues can elicit para-social relationships, where users respond to media figures as if they were real \cite{horton_mass_1956,nass_machines_2000}. Relational agents demonstrate this effect: consistent use of social cues such as empathy, self-disclosure, or small talk helps sustain trust and engagement over extended interactions \cite{bickmore_establishing_2005}. More recent work refines these insights by showing that users value specific types of cues differently (e.g., empowerment, affirmation, alliance-building, or social dialogue) when relating to digital coaches \cite{salman_identifying_2023}. Meta-analytic evidence confirms that human-like cues, while subtle, reliably enhance perceptions of presence, rapport, and trust in conversational agents \cite{klein_effects_2025}. Complementing this, affiliative humor has been shown to increase motivation, build rapport, and make health counseling conversations with agents more engaging \cite{olafsson_motivating_2020}. Extending these findings to online fitness, interactivity, authenticity, and companionship have been identified as critical drivers of para-social relationships, with emotional support proving more effective than purely technical guidance \cite{feng_virtual_2025}. Recent studies further show that agents perceived as socially present can sustain motivation even when users are fully aware of their artificial nature \cite{oh_enhancing_2025}. While these relational mechanisms highlight the promise of AI peers to foster presence and motivation, they also expose a deeper fragility: social presence can be generated, yet may not always be accepted as authentic.

\par Yet, a critical tension remains. Even when AI peers are designed to be believable, human-like, and socially present, users may still resist accepting them as genuine companions. People often value agents for their availability and non-judgmental support, but dismiss them as lacking the capacity for mutual caring -- a quality seen as essential for ``real'' relationships \cite{oguz-uguralp_why_2025}. Beyond authenticity, safety concerns also emerge: generative AI wellness applications can pose health risks when agents mishandle sensitive disclosures or blur the boundaries between coaching and medical advice \cite{de_freitas_health_2024}. In physical activity specifically, recent studies show that users remain skeptical of simulated peers unless they display visible and believable signs of exertion; otherwise their performance is dismissed as “just a number,” undermining trust and motivation \cite{silacci_when_2025}. Together, these findings underscore the fragility of artificial companionship: social presence can be simulated, but authenticity, safety, and sustained motivational depth remain open challenges.

\par This work addresses that fragility by examining whether AI peers can motivate sustained physical activity, and under what relational conditions they succeed. Moving beyond the dominant coach model, we design and evaluate LLM-powered peers that act as exertion partners rather than authority figures. Our \revision{four-group randomized controlled trial} study empirically tests how such peers compare to human counterparts in fostering motivation, social presence, and engagement over time. By situating AI peers within real PA contexts, we contribute evidence on both their potential and their limits, highlighting design strategies that can enhance believability, humanness, and presence while acknowledging the challenges of authenticity and safety.

\subsection{Research questions and hypotheses}
\label{sub:rw_rqs_hyps}
Our review of the literature highlights the powerful role of social support in promoting physical activity, but also reveals a critical tension in how this support is best delivered via technology. While human peers provide authentic accountability, AI agents offer a scalable and consistent source of motivation. This leads to fundamental questions about the comparative effectiveness of human versus AI-driven support, how an agent's embodiment influences the user's experience, and how users perceive these novel social partners. To investigate these issues, our work is guided by the following research questions and hypotheses:
\begin{itemize}
    \item \textbf{RQ1: To what extent can AI agents effectively provide relatedness support to promote and sustain physical activity?} Grounded in Self-Determination Theory and prior work demonstrating that social influence fosters engagement and motivation, we predict that having a peer -- either human or AI -- will be more effective than exercising alone. We also explore the stability of this behavior change, positing that the consistent nature of an AI peer may lead to more durable habits compared to the potentially transient motivation from a human partner.
    \begin{itemize}
        \item \textbf{H1a} The increase in monthly steps from the baseline to the deployment phase will be greater in the Human-Human and Human-SEP conditions than in the Control condition.
        \item \textbf{H1b} The decrease in monthly steps from the deployment phase to the post-deployment phase will be smaller in the Human-SEP and Control conditions than in the Human-Human condition.
        \item \textbf{H1c} The overall increase in monthly steps from the baseline to the post-deployment phase will be greater in the Human-Human and Human-SEP conditions than in the Control condition.
    \end{itemize}
    \item \textbf{RQ2: How do AI agents compare to humans in providing relatedness support to promote and sustain physical activity?} While human relationships can be powerful, their effectiveness is often variable; shyness or a lack of shared values can prevent a strong bond from forming. In contrast, AI agents can provide consistently available and non-judgmental support, which may foster a stronger sense of psychological safety and perceived relatedness. Therefore, we posit:
    \begin{itemize}
        \item \textbf{H2a} During the deployment phase, participants in the Human-SEP conditions will report higher perceived relatedness (IMI) than participants in the Human-Human condition.
        \item \textbf{H2b} During the deployment phase, participants in the Human-SEP conditions will report higher perceived competence (IMI) than participants in the Human-Human condition.
        \item \textbf{H2c} During the deployment phase, participants in the Human-SEP conditions will report higher social presence (SPS) than participants in the Human-Human condition.
        \item \textbf{H2d} During the deployment phase, participants in the Human-SEP conditions will form a stronger working alliance bond (WAI) than participants in the Human-Human condition.
    \end{itemize}
    \item \textbf{RQ3: How does the visual appearance of an AI agent, specifically a human figure versus a cyborg, affect its ability to provide relatedness support?} Building on work showing that anthropomorphic features can increase trust and satisfaction, we explore how a human-like versus a non-human appearance affects the user's connection with an AI peer.
    \begin{itemize}
        \item \textbf{H3a} During the deployment phase, participants in the Human-SEP\_human condition will report higher perceived relatedness (IMI) than those in the Human-SEP\_cyborg condition.
    \end{itemize}
    \item \textbf{RQ4: How do users actively assess and evaluate the virtual peers' visual appearance, behavior, and communication style, considering both their level of believability and the potential for perceived deception within the context of a physical activity intervention?} This final research question is exploratory and qualitative. It aims to understand the mechanisms behind our quantitative findings by examining how users form judgments about their peers. We seek to understand how coherence, communication style, and behavior influence the believability of an agent and the user's overall experience, a critical factor given that a mismatch between expectations and agent behavior can lead to frustration. 
\end{itemize}
\section{Methods}
\label{sec:methods}
The nature of our research questions, which focus on sustained behavior change and relational dynamics, necessitated a longitudinal approach. We conducted a six-month, four-group, randomized controlled trial to investigate the effectiveness of AI-driven virtual agents in providing social support for physical activity. This long-term design was crucial as it enabled us to move beyond a single snapshot in time. This was necessary because behavior change is not a discrete event but a complex process that unfolds over a person's life course, with technology use evolving as life circumstances change \cite{rapp_exploring_2023}. Furthermore, a shorter study would have been insufficient to mitigate the initial novelty effect; prior research has shown that users' initial curiosity with new health technologies can fade quickly, leading to high rates of abandonment if the tool's long-term value is not established \cite{rapp_personal_2016}. Additionally, given our focus on relational constructs like working alliance and social presence, the extended duration was necessary to capture how these dynamics evolved \cite{bickmore_establishing_2005}. Finally, our design included a three-month post-intervention phase to assess the persistence of any behavior change after the social support was removed. This is critical because prior longitudinal research has shown that some persuasive techniques can undermine a user's intrinsic motivation, leading to a decrease in the target behavior once the intervention is withdrawn \cite{cherubini_unexpected_2020}. Our follow-up phase was therefore essential to evaluate the intervention's lasting behavioral impact.

\par To ensure transparency and support reproducibility, this study adheres to the guidelines proposed by Salehzadeh Niksirat et al. \cite{salehzadeh_niksirat_changes_2023}. All research materials are available in a dedicated OSF repository\footnote{See: \url{https://osf.io/k3gvh/overview?view_only=0a2b40f80e5d4ec9b3fb96838875e3f1}, last accessed January 2026}. The repository includes: (1) all quantitative data and analysis scripts; (2) the full qualitative pipeline, including the interview guide, codebook, and final themes; and (3) supplementary materials such as the translated quotes used in this paper.

\subsection{Participants and recruitment}
\label{sub:methods_participants_and_recruitment}
A total of 280 participants were recruited from the \anonymize{University of Lausanne's ORSEE} volunteer pool, which consists of approximately 8,000 students. This target sample size was determined by the practical constraints of the available research budget. An a-priori power analysis indicated that this sample size would provide a power of .62 to detect medium-to-large effect sizes. While this is below the conventional 0.80 threshold, the study is adequately powered for robust effects. However, it may be underpowered to reliably detect smaller effects. Consequently, while the significant findings reported in this paper are likely robust, any non-significant results should be interpreted with caution.

\par Eligibility criteria required participants to be at least 18 years of age, fluent in French, own an Apple iPhone running iOS 8 or later, and provide consent for two key processes: the collection of their daily step count data via Apple's HealthKit framework, and the use of third-party services, namely OpenAI's LLM API and Google's Firebase Cloud Messaging, for the intervention's functionality. A screening process, utilizing the International Physical Activity Questionnaire French short form (IPAQ-SF) \cite{craig_international_2003} and the Health Behavior and Stages of Change Questionnaire (HBSCQ) \cite{gonzalez-ramirez_validation_2017}, was employed to select individuals who exhibited at least a moderate level of physical activity and were identified as being at least in the "Preparation" stage of behavior change -- a stage where individuals recognize the need for change and ready to take action \cite{prochaska_transtheoretical_1997}. Our recruitment efforts aimed to maintain a balanced gender distribution across the sample.

\par For the post-study interviews, we purposefully selected 30 participants from the main study pool. This selection was primarily guided by their responses to the three questionnaires administered during the deployment phase to ensure a range of experiences, and secondarily by their availability. We made a best effort to ensure balanced representation from the HUM, SEPC, and SEPH conditions and across genders.

\subsection{Experimental design}
\label{sub:methods_experimental_design}
The study employed a between-subjects design, with participants randomly assigned to one of four conditions: (1) a Control (CON) condition, where participants exercised individually; (2) a Human-Human (HUM) condition, where participants were paired with another \revision{participant}; (3) a Human-$SEP_{human}$ (SEPH) condition, where participants were paired with a SEP represented by a human-like avatar; and (4) a Human-$SEP_{cyborg}$ (SEPC) condition, where participants were paired with a SEP represented by a cyborg avatar.

\par In all conditions, participants were given a weekly step count goal of 70000 steps (10000 steps per day). In the three dyadic conditions (HUM, SEPH, and SEPC), participants could view their peer's progress toward the daily goal -- without seeing the peer's exact number of steps -- and communicate through an integrated chat feature. This design allowed for the direct comparison of social support provided by a human peer versus two distinct types of peers. Participants in the Control condition received the same step goals but had no peer interaction.

\subsection{Apparatus}
\label{sub:methods_apparatus}
\begin{figure*}[ht]
    \centering
    \includegraphics[width=.8\textwidth]{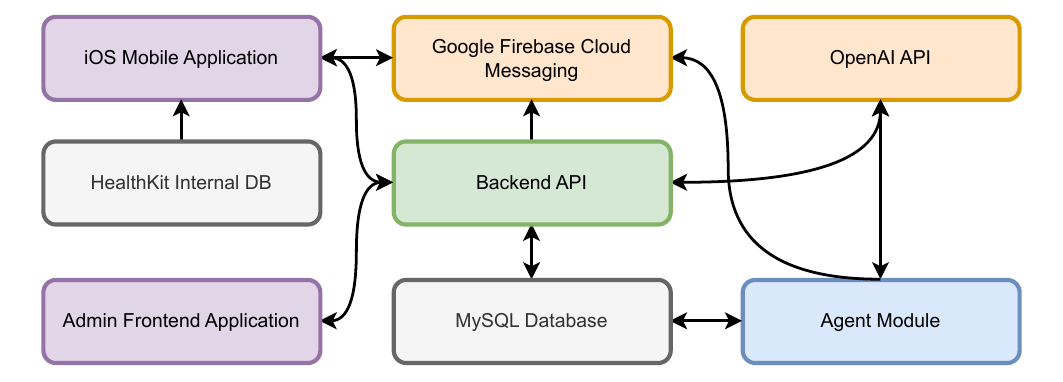}
    \caption{The architecture of the intervention illustrating the communication and directionality of the content between the different components.}
    \Description{System architecture of the intervention. The diagram shows multiple components connected through a central Backend API. On the user side, an iOS mobile application and Apple HealthKit database communicate with the backend. On the researcher side, an admin frontend application and a MySQL database connect to the backend. External services include Google Firebase Cloud Messaging for notifications, the OpenAI API for language generation, and an agent module representing the AI peer.}
    \label{fig:architecture}
\end{figure*}

This study relied on two important components, the first being the mobile intervention, \anonymize{Excero}, and the integration of Alex, the AI-driven agent for the SEPC and SEPH conditions. The entire system was supported by a custom backend for data management, API integrations with OpenAI and Firebase, and a researcher-facing dashboard for monitoring the study's progress. We detail the architecture in Figure \ref{fig:architecture}, and provide more insights on the features that concern the two main components.

\subsubsection{Excero: the mobile application}
\label{subs:methods_apparatus_excero}
\begin{figure*}[ht]
    \centering
    \begin{subfigure}[t]{.271\textwidth}
        \includegraphics[width=\textwidth]{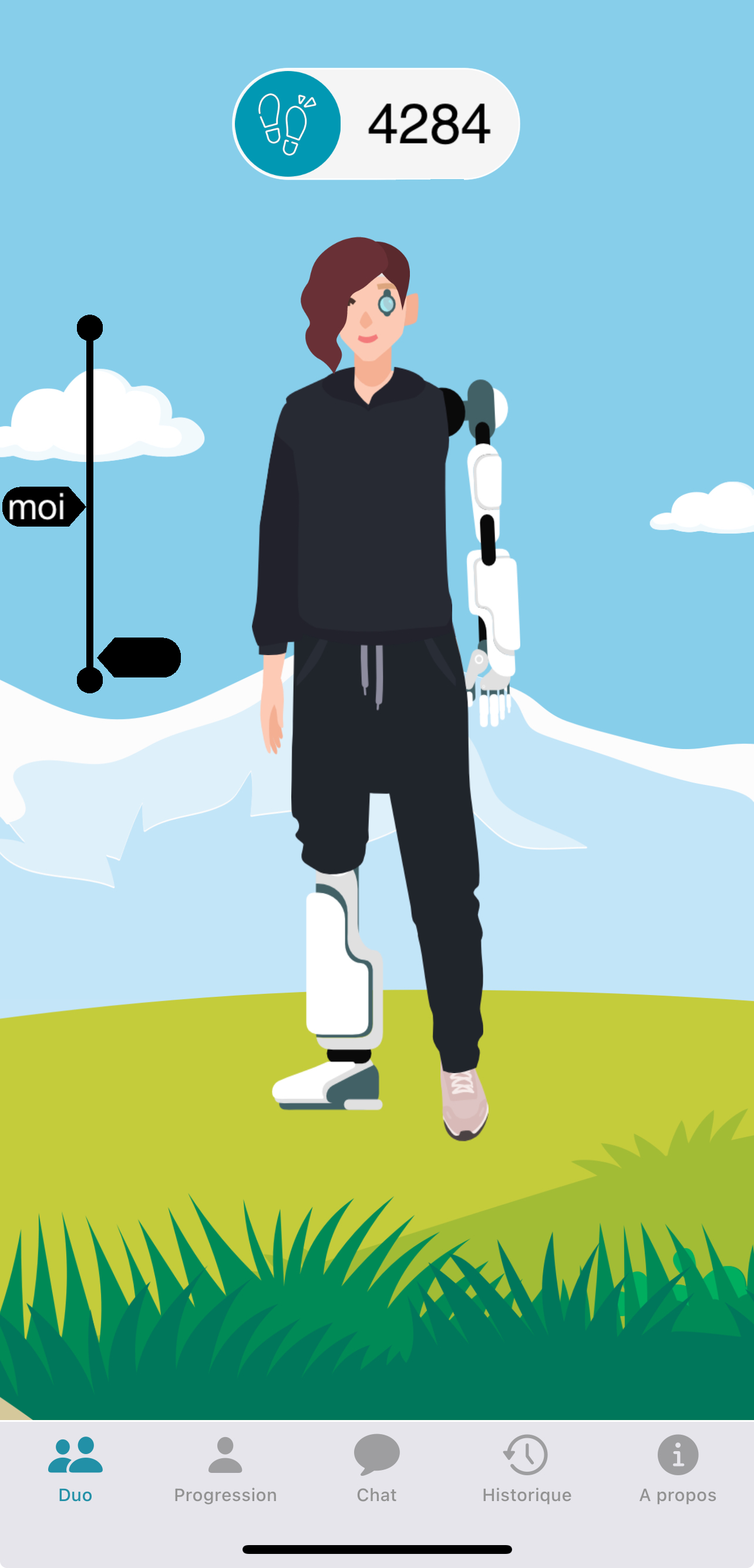}
        \caption{Duo view of the Simulated Exercising Peer with a combined human-robot appearance (SEPC condition). Participants viewed their own step count above the agent and could compare progress toward the daily 10,000-step goal ("moi" meaning "me" in French, to indicate that it represents the user's progress).}
        \Description{Duo view of the mobile app in the SEPC (Simulated Exercising Peer - cyborg) condition. The screen shows a humanoid avatar with robotic limbs representing the AI peer, alongside the user’s progress bar ("moi": "me" in French) and the user's step count of 4,284 displayed at the top. On the bottom are navigation icons for progression, chat, history, and information are visible at the bottom.}
        \label{fig:screenshots_sepc_duo}
    \end{subfigure}
    \begin{subfigure}[t]{.272\textwidth}
        \includegraphics[width=\textwidth]{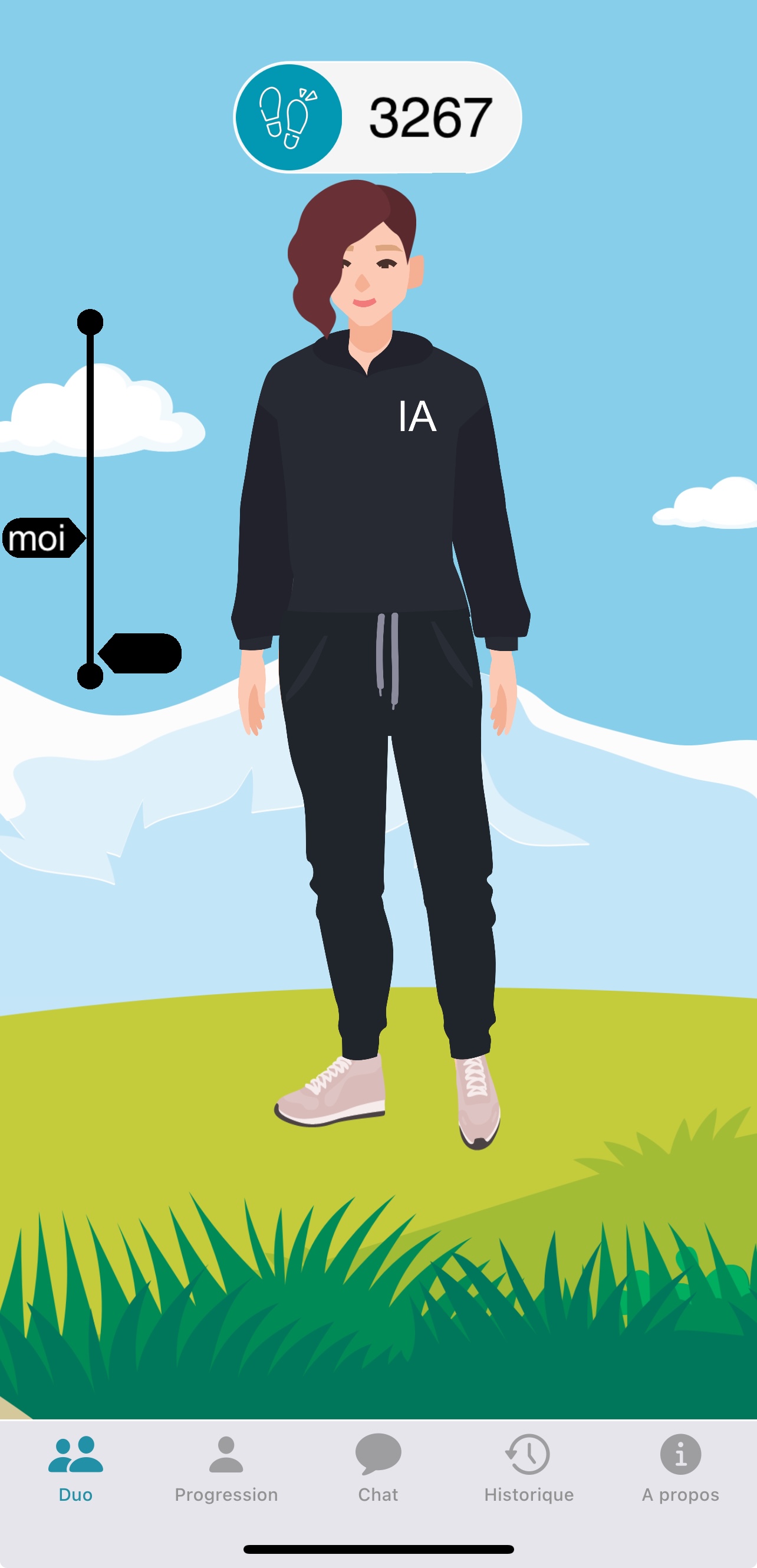}
        \caption{Duo view of the Simulated Exercising Peer with a fully human appearance and an ``IA'' label on the torso (SEPH condition). Participants viewed their own step count above the agent and compared progress toward the daily 10,000-step goal ("moi" meaning "me" in French, to indicate that it represents the user's progress).}
        \Description{Duo view of the mobile app in the SEPH (Simulated Exercising Peer - human) condition. The screen shows a human-like avatar wearing a black hoodie labeled "IA," representing the AI peer. The user’s progress bar ("moi": "me" in French) is displayed on the left, and the user’s step count of 3,267 appears at the top. On the bottom are navigation icons for progression, chat, history, and information are visible at the bottom.}
        \label{fig:screenshots_seph_duo}
    \end{subfigure}
    \begin{subfigure}[t]{.27\textwidth}
        \includegraphics[width=\textwidth]{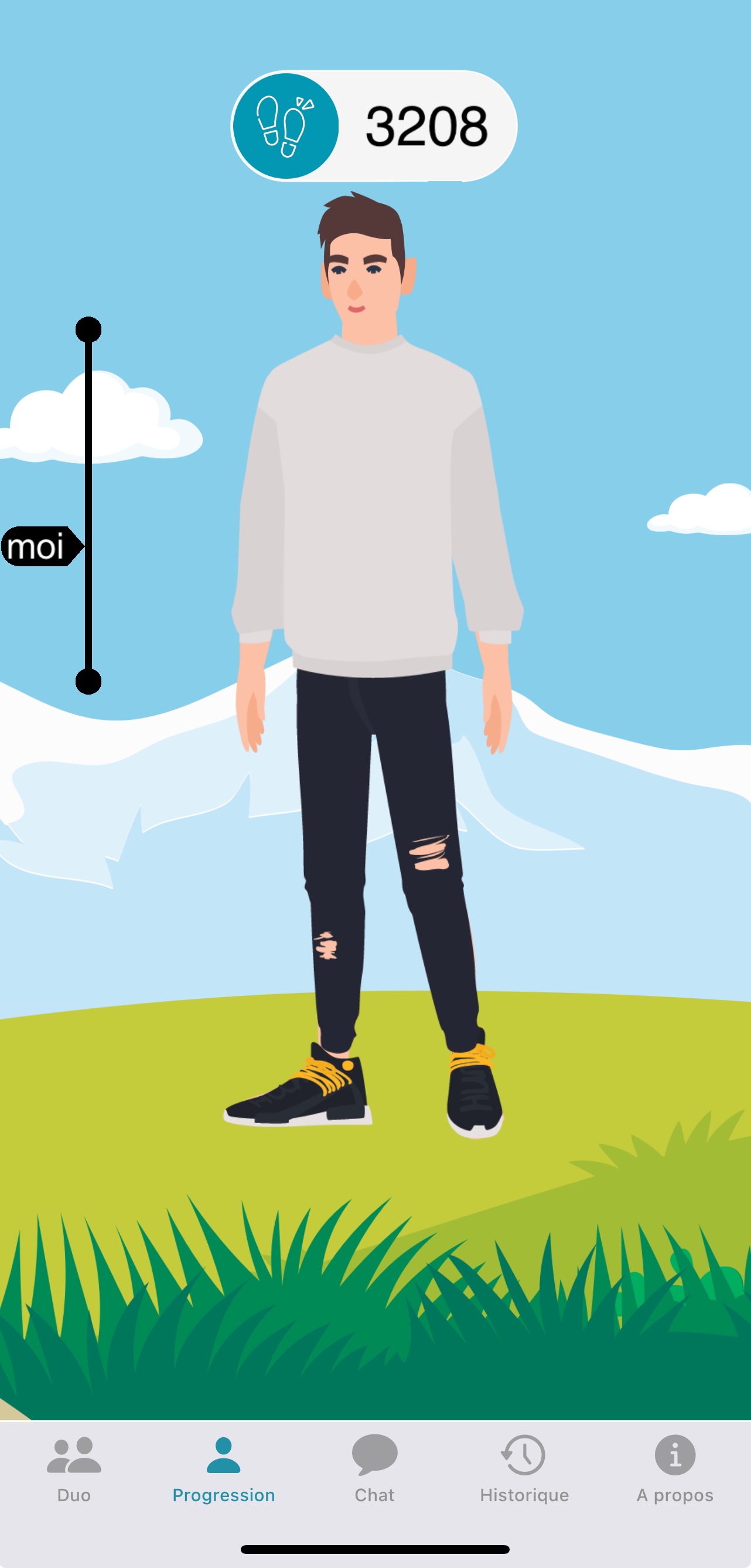}
        \caption{Progression view showing the participant’s avatar with their step count displayed above it. Participants could compare their own progress with the agent’s toward the daily 10,000-step goal. This view was available in all conditions ("moi" meaning "me" in French, to indicate that it represents the user's progress).}
        \Description{Progression view of the mobile app. The screen shows a human avatar standing outdoors, representing the user’s own character. A vertical bar on the left labeled "moi" indicates the user’s progress, with the user's step count (3,208) displayed at the top. On the bottom are navigation icons for duo, progression, chat, history, and information are visible at the bottom.}
        \label{fig:screenshots_progression}
    \end{subfigure}
    \caption{Screenshots of the Duo views (Simulated Exercising Peer as a cyborg in SEPC, and as a human in SEPH) and the Progression view from the Excero mobile application.}
    \Description{Screenshots of the mobile application showing the different peer conditions and user progress view. The Duo view displays the user alongside a simulated exercising peer: (left) SEPC condition with a humanoid avatar featuring robotic limbs, and (center) SEPH condition with a fully human-like avatar labeled "IA." The Progression view (right) focuses on the user’s own avatar and step count without showing a peer. In all views, step counts are displayed at the top, a vertical bar on the left marks the user’s progress ("moi": "me" in French), and navigation icons appear at the bottom.}
    \label{fig:screenshots_part_1}
\end{figure*}

The intervention was delivered through \anonymize{Excero}, an iOS mobile application developed for this study (see Figure \ref{fig:screenshots_part_1} and Figure \ref{fig:screenshots_part_2}). The application provided several key features. Upon joining, users could personalize a cartoon-style avatar. The primary interface consisted of a ``Progression'' view, as illustrated in Figure \ref{fig:screenshots_progression}, which displayed the user's avatar and a visual indicator of their progress toward their weekly step goal. In the dyadic conditions, an additional ``Duo'' view presented the peer's avatar and their corresponding progress (e.g., Figure \ref{fig:screenshots_sepc_duo} and Figure \ref{fig:screenshots_seph_duo}). Figure \ref{fig:screenshots_history} illustrates the ``History'' view that allowed participants to review their daily step counts and track their weekly goal achievements. A text-based chat interface was available in the dyadic conditions, enabling communication between peers and providing a feature to report inappropriate behavior. Users received push notifications to alert them of new messages or to remind them to synchronize their step data if they had not opened the application in seven days (see Figure \ref{fig:screenshots_chat}). Finally, a specific view was used to inform participants on the synchronization status of their steps with our backend and mentioned that the experiment was still ongoing (see Figure \ref{fig:screenshots_sync}).

\subsubsection{Alex: the simulated exercising peer}
\label{subs:methods_apparatus_alex}
In the SEP conditions, participants were paired with an AI-driven peer named Alex, who was also represented by an avatar. \revision{In the SEPH condition, Alex was presented as a human wearing a hoodie with an "AI" logo to clearly signal its non-human nature (see Figure \ref{fig:screenshots_seph_duo}). In the SEPC condition, Alex was depicted as a cyborg, with robotic features (see Figure \ref{fig:screenshots_sepc_duo}).} 

\par \revision{We use the term cyborg following Haraway’s definition of a cybernetic organism as a hybrid of machine and organism \cite{weiss_cyborg_2006}: Haraway does not specify whether the dominant substrate is biological or mechanical, so a cyborg may originally be conceived either as a machine with some organic parts (like skin) or as a human with some mechanical parts (like a prosthetic arm). In our design, ``cyborg'' refers to an artificial system with partially human-like features -- akin to science-fiction representations such as the T-800 Terminator -- which appear outwardly human while revealing their artificiality. This framing aligns with our goal of creating an avatar that felt closer to human than a robot, yet still unmistakably artificial so as not to mislead the user. A robot-like avatar, on the other hand, would not have conveyed the human appearance of the artificial agent. Importantly, this terminology was never shown to participants, and Alex consistently presented itself as an AI agent in all interactions, preventing any risk of misinterpretation of its ontological status.}

\par Alex's conversational responses were generated using OpenAI's GPT-4o model. The agent was programmed to be both reactive and proactive. It responded whenever a user initiated a conversation, and it would proactively contact the participant at 1:00 PM if no communication had occurred in the previous 24 hours. For proactive messages, the LLM prompt included the participant's username, the current day, and their daily and weekly step progress to generate a contextually relevant message.

\par To foster a sense of social comparison, Alex was also designed to log its own daily steps. The number of steps Alex would ``achieve'' each hour was dynamically generated by OpenAI's GPT-4o model. \revision{Thus, both the conversational and behavioral components relied on the same version of the model}. The prompt for this function included the participant's step count from the previous hour and their total steps from the same day in the previous week. \revision{Based on the users' steps data, the} LLM was instructed to generate a step count that would create a sense of optimal challenge for the user, aiming to encourage a healthy, non-discouraging social comparison. \revision{The complete set of prompts used in this study is included in the \href{https://osf.io/k3gvh/overview?view_only=0a2b40f80e5d4ec9b3fb96838875e3f1}{Supplementary Material}.}

\par
\begin{figure*}[ht]
    \begin{subfigure}[t]{.27\textwidth}
        \includegraphics[width=\textwidth]{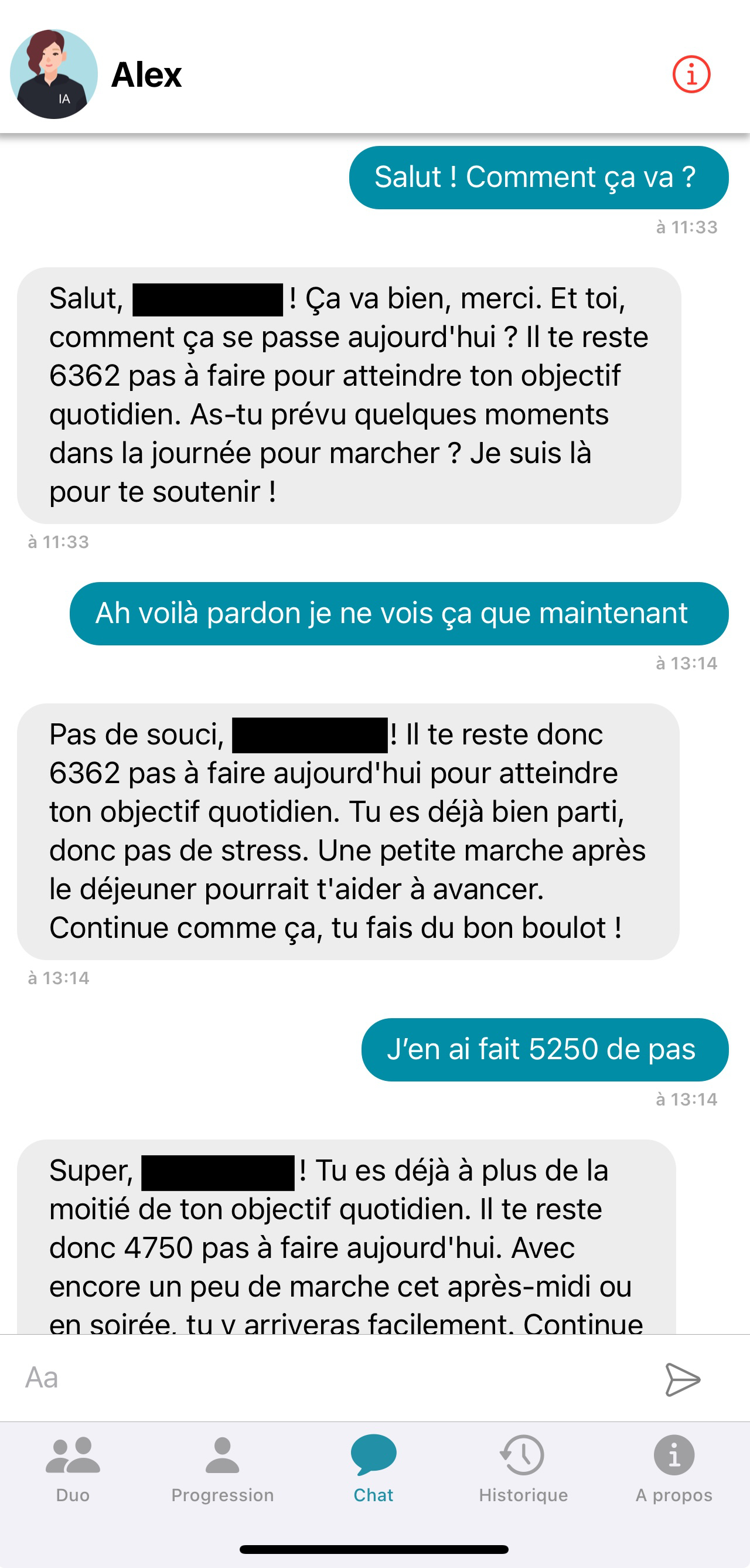}
        \caption{Chat view used in the SEPC, SEPH, and HUM conditions. Participants could exchange free-text messages (including emojis) with their peer, whose profile picture and username were displayed. A red icon allowed participants to report their peer in case of discomfort.}
        \Description{Chat view of the mobile app, used in the SEPC, SEPH, and HUM conditions. The screenshot shows a conversation between the user and the peer (here labeled "Alex", the SEP), where the peer provides encouragement and feedback on step counts in French. The messages highlight remaining steps toward the daily goal, suggest walking strategies, and reinforce progress. On the bottom are navigation icons for duo, progression, chat, history, and information are visible at the bottom.}
        \label{fig:screenshots_chat}
    \end{subfigure}
    \begin{subfigure}[t]{.27\textwidth}
        \includegraphics[width=\textwidth]{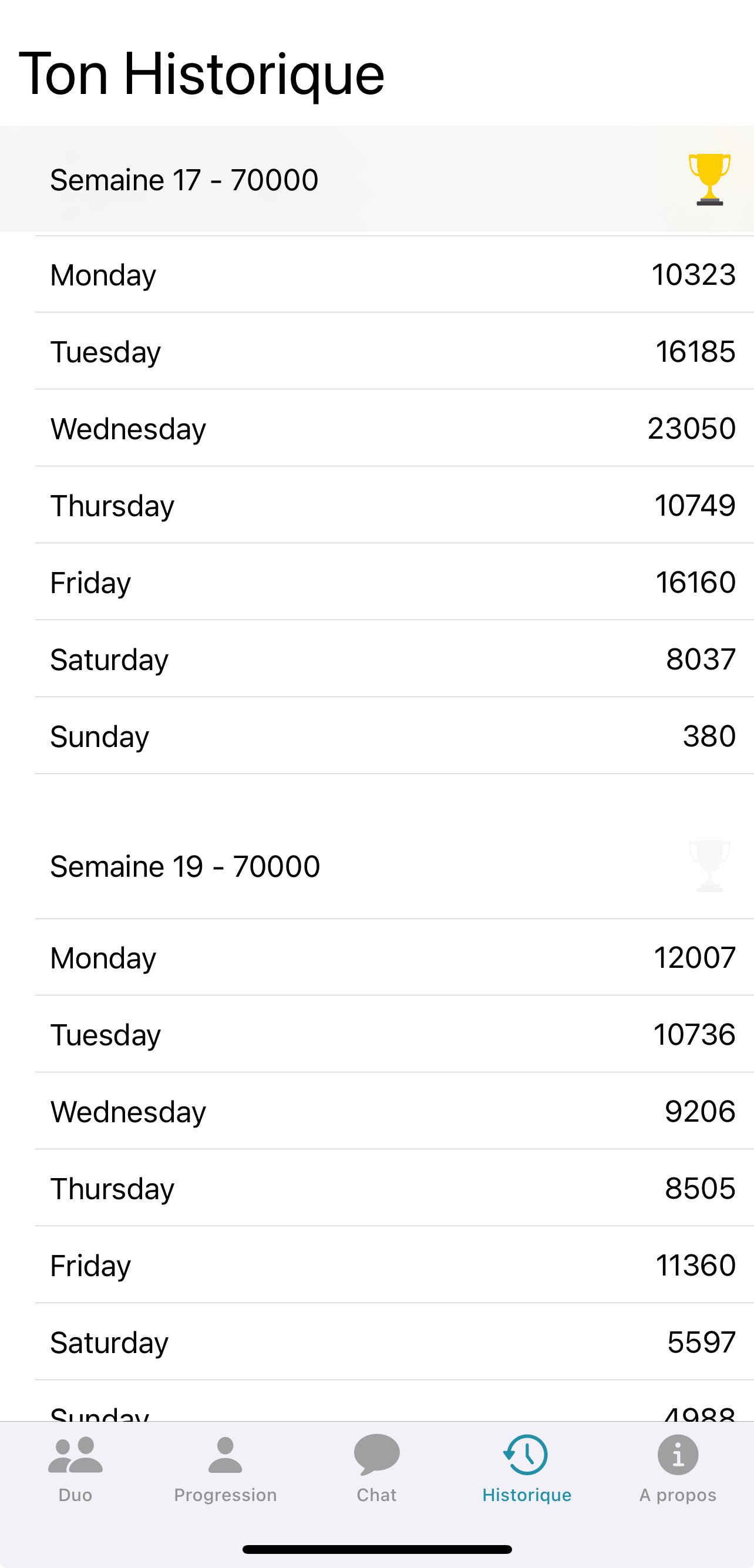}
        \caption{History view displaying daily step counts collected from HealthKit in a weekly list. Each header shows the week number, the 70,000-step goal, and a colored trophy if the goal was met (gray if not). This view was available in all conditions.}
        \Description{History view of the mobile app. The screen lists daily step counts for two weeks, grouped under weekly goals of 70,000 steps. Each day (Monday to Sunday) shows the number of steps completed, with a trophy icon marking weeks where the goal was reached. On the bottom are navigation icons for duo, progression, chat, history, and information are displayed at the bottom.}
        \label{fig:screenshots_history}
    \end{subfigure}
    \begin{subfigure}[t]{.27\textwidth}
        \includegraphics[width=\textwidth]{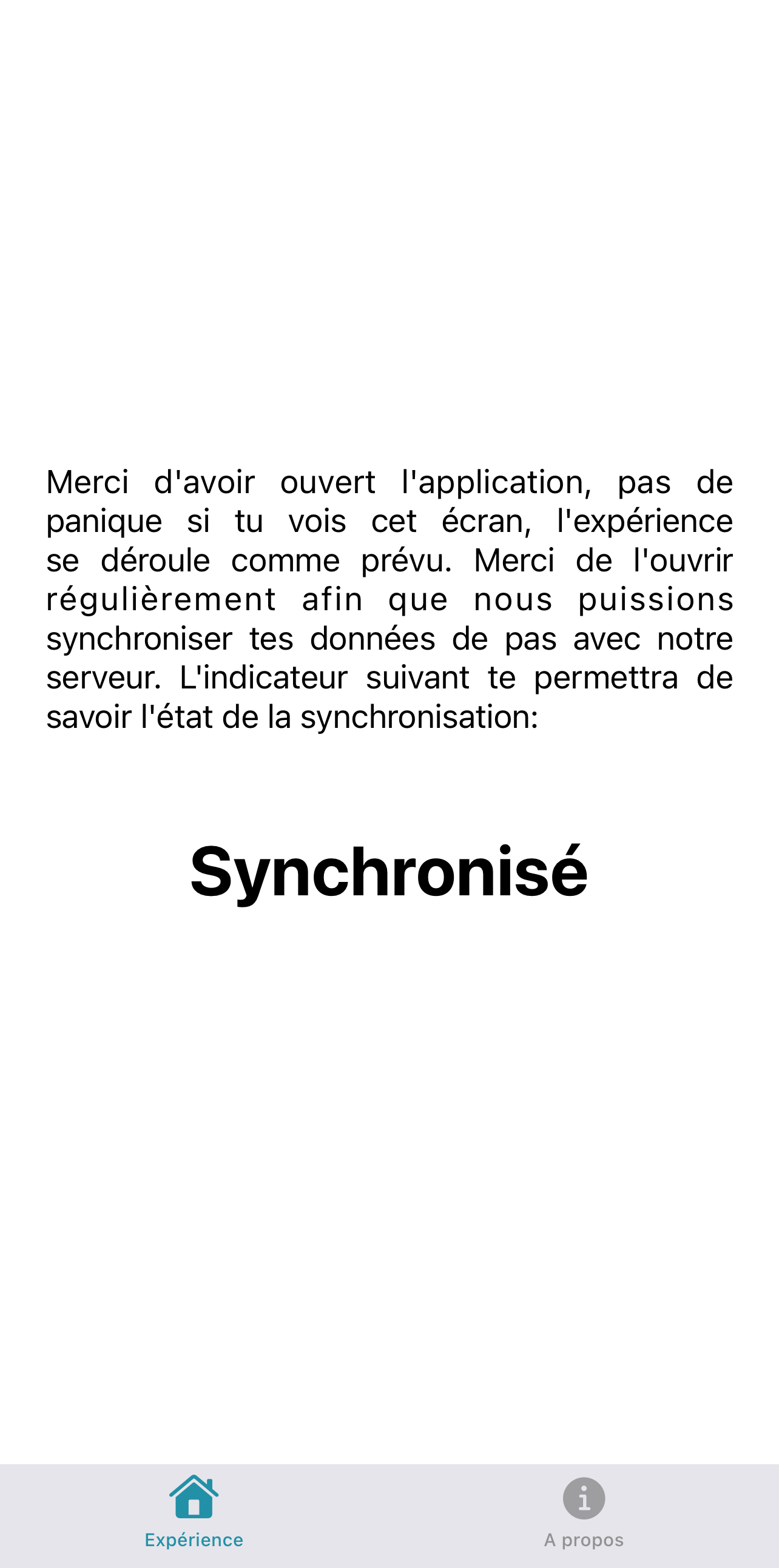}
        \caption{View displayed outside the deployment phase. Participants were informed that the experiment was ongoing and could monitor the synchronization status of their step data with the app’s backend.}
        \Description{Synchronization view of the mobile app. The screen displays a confirmation message in French reassuring participants that the study is proceeding as planned and reminding them to regularly open the app to sync their step data with the server. A large label reads "Synchronisé" to indicate that synchronization is active. On the bottom are navigation icons for experience and information are visible at the bottom.}
        \label{fig:screenshots_sync}
    \end{subfigure}
    \caption{Screenshots of the Chat view (SEPC, SEPH, and HUM conditions), the History view, and the default view displayed outside the deployment phase in the Excero mobile application.}
    \Description{Screenshots of the mobile application highlighting key interface views beyond the Duo and Progression screens. The Chat view (left) shows a text-based conversation between the user and the peer (AI or human), in the example, a SEP provides motivational feedback and updates on step goals. The History view (center) lists daily step counts grouped by week, with icons indicating whether weekly goals were achieved. The Synchronization view (right) reassures participants that the study is ongoing outside the active deployment window and confirms that data are being synced, with the large status label "Synchronisé" displayed.}
    \label{fig:screenshots_part_2}
\end{figure*}

 \revision{Participants were informed from the outset that Alex was an AI agent. They were provided with multiple channels to request support or flag concerns about the agent’s behavior, including email contact and an in-app report button in the chat interface (cf., Figure \ref{fig:screenshots_chat}). To ensure participant safety, researchers conducted weekly reviews of all chat transcripts to verify that Alex’s responses remained accurate, non-harmful, and appropriate. Any messages containing potentially misleading or unsafe physical-activity advice were flagged for follow-up and participants were contacted when necessary.}

\subsection{Procedure}
\label{sub:methods_procedure}
The experiment began with a two-week Onboarding phase. After passing the screening questionnaire, selected participants received an email with a link to the informed consent form and a request to choose a username. Upon consenting, they were randomly assigned to a condition and received a second email with instructions for downloading and logging into the mobile application. \revision{Participants were assigned to one of the four experimental conditions using a blocked and stratified randomization procedure. Assignment was blocked to maintain approximately equal numbers of participants per condition, and stratified by self-identified gender to ensure comparable gender distributions across conditions.} This two-week period ensured all participants could successfully install the app and allowed for the collection of baseline step data via Apple HealthKit. During this phase, the app displayed a screen confirming that the experiment was being set up and that their data was being synchronized.

\par The three-month Deployment phase began on October 28th, 2024, and concluded on January 28th, 2025. This timeframe was chosen to capture student behavior across different periods of the academic year (the main semester, winter holidays, and the exam period). During this phase, participants had access to all features of the intervention corresponding to their assigned condition. The app synchronized their step history each time it was opened. On the 28th of each month, participants received an email invitation to complete a questionnaire (T1: Nov 28, T2: Dec 28, T3: Jan 28).

\par In the final three-month Post-Deployment phase, all social and interactive features of the application were disabled. However, the app continued to collect daily step counts to assess any lasting effects of the intervention on physical activity. After the study's conclusion, we conducted semi-structured interviews with a representative subset of participants to gather qualitative data about their experiences.

\par Participants who completed all questionnaires and maintained regular data synchronization received a total of CHF 60 ($\approx$ USD 75). This was paid in two installments: CHF 40 ($\approx$ USD 50) after the deployment phase and CHF 20 ($\approx$ USD 25) at the end of the post-deployment phase. Interviewees received an additional CHF 20. Participants who failed to complete questionnaires but continued to synchronize data received a partial payment of CHF 20. Participants could be excluded without payment for failing to complete any tasks or for being reported for inappropriate communication (in the HUM condition), pending researcher review. All participants were informed they could withdraw at any time and request the complete removal of their data. 

\subsection{Measures}
\label{sub:methods_measures}
Our analysis relied on a combination of quantitative and qualitative data, all collected in French.

\subsubsection{Physical activity}
The primary outcome measure for physical activity was the daily step count, collected objectively and automatically via Apple's HealthKit framework on the participants' iPhones.

\subsubsection{Psychological experience}
\label{subs:psychological_experience}
To assess participants' psychological experience, we administered several validated scales. As our study population was French-speaking, all scales underwent a rigorous translation process from English to French. This process involved two professional translators (C2 level according to the Common European Framework of Reference for Languages), with a third external expert resolving any discrepancies. A final verification for accuracy and cultural appropriateness was then conducted by a member of the research team fluent in French. All items were rated on a 7-point Likert scale ranging from 1 (Strongly Disagree) to 7 (Strongly Agree). \revision{We selected these validated instruments because they could be applied consistently across both human-human and human-AI pairings and were available in validated French versions at the time of data collection, ensuring conceptual alignment and comparability across conditions.}
\begin{itemize}
\item \textbf{Intrinsic Motivation Inventory (IMI).} We used two subscales from the IMI \cite{deci_intrinsic_1985} to measure perceived relatedness (e.g., \textit{``I felt really distant to this person.''}) and perceived competence (e.g., \textit{``I think I am pretty good at this activity.''}).
\item \textbf{Social Presence Scale (SPS).} To evaluate the sense of connection with the peer, we used the SPS \cite{gefen_consumer_2004}. A sample item is \textit{``There is a sense of sociability with Alex, the chatbot.''}
\item \textbf{Working Alliance Inventory (WAI).} We used the bond subscale from the WAI \cite{horvath_development_1989} to assess the quality of the collaborative bond. A sample item is \textit{``I feel that Alex appreciates me.''}
\end{itemize}

\subsubsection{Qualitative data}
Qualitative data were derived from the thematic analysis of transcripts from the 30 post-deployment semi-structured interviews.

\subsection{Data analysis}
\label{sub:methods_data_analysis}
\subsubsection{Quantitative analysis}
\label{subs:methods_data_analysis_quantitative}
For the primary outcome of daily step counts, we fit a linear mixed-effects model (LMER) with participants as a random effect to account for repeated measures. Fixed effects were evaluated both at the level of model coefficients ($\beta$, SE, t, p, CI) and at the factor level using omnibus likelihood ratio tests ($\chi^2$). Coefficients are reported in regression tables, while full omnibus test results are provided in the \href{https://osf.io/k3gvh/overview?view_only=0a2b40f80e5d4ec9b3fb96838875e3f1}{Supplementary Material}. We used Estimated Marginal Means (emmeans) with a Tukey correction for post-hoc pairwise comparisons.

\par For the quantitative questionnaire data, we conducted Mixed ANOVAs. Time (measurements at T1, T2, and T3) was treated as a within-subjects factor, while Condition was treated as a between-subjects factor. This model allowed us to examine the main effects of time and condition, and critically, their interaction. Significant effects were explored further using pairwise t-tests with a Bonferroni correction for multiple comparisons. All quantitative analyses were conducted using R.

\subsubsection{Qualitative analysis}
\label{subs:methods_data_analysis_qualitative}
We analyzed the interview transcripts using a \revision{coding reliability} thematic analysis approach\revision{, consistent with post-positivist small-q forms of thematic analysis and in line with Braun and Clarke’s \cite{braun_toward_2023} discussion of coding-reliability methodologies}. The analysis began with two researchers collaboratively developing a preliminary codebook based on one interview from each of the three dyadic conditions (HUM, SEPC, and SEPH). To ensure the reliability of the coding process, the two researchers then independently coded 10\% of the transcripts. Following an initial round where agreement was insufficient, the codebook was refined through discussion. A second round of independent coding on a new subset of data achieved a Cohen's Kappa of 0.81, indicating substantial inter-rater reliability \cite{mchugh_interrater_2012}. One researcher then coded the remaining data using the established codebook. The final stage of analysis involved an iterative process of theme development, where the initial set of codes was refined and consolidated into 41 final codes, then organized into 7 sub-themes and, finally, 3 overarching themes\revision{: Perceptual assessment of virtual peer, Believability and deception, and Relational dynamics with peers}.

\subsection{Ethics}
\label{sub:methods_ethics}
The study protocols for both the six-month study and the interviews received approval from the Internal Review Board (IRB) of the \anonymize{Faculty of Business and Economics} at the \anonymize{University of Lausanne}. All participants provided written informed consent prior to the start of the study and again before the interviews. Participants were informed that their participation was voluntary and that they could withdraw at any time without penalty. To protect privacy, participant data on our backend database was pseudonymized, and user IDs were encrypted when using third-party services. All collected data, including step counts and conversation logs, were fully anonymized before analysis. The data analysis was performed on blinded condition information, which was unblinded only at the final stage of the process to prevent any potential bias. 
\section{Results}
\label{sec:results}
Of the 280 participants who were recruited and randomized, 28 failed to complete the onboarding process (8 withdrew after enrollment; 20 never activated their account). This left a starting sample of 252 participants. Of these, 250 synchronized their step data for at least a portion of the study (120 female, 77 male, 2 other/non-binary), while 128 participants completed the full six-month data synchronization \revision{(68 female, 58 male, 2 other/non-binary; mean age = 21.2, SD = 2.66, see \href{https://osf.io/k3gvh/overview?view_only=0a2b40f80e5d4ec9b3fb96838875e3f1}{Supplementary Material} for detailed demographics)}, resulting in a final attrition rate of 54\% for the primary behavioral measure. For the primary outcome of daily step counts, all 128 participants who provided some data were included in the Linear Mixed-Effects Model (LMER) analysis, as the model can accommodate missing daily data points.

\par For the analysis of the questionnaire data (IMI, SPS, and WAI), the sample size varied at each time point due to non-completion. At T1, 218 participants completed the questionnaire; at T2, 204 completed it; and at T3, 193 completed it. The Mixed ANOVA analyses were conducted on the \revision{141 participants (79 female, 61 male, 1 other/non-binary; mean age = 21.6, SD = 2.55)} who provided complete data for all three time points \revision{(see \href{https://osf.io/k3gvh/overview?view_only=0a2b40f80e5d4ec9b3fb96838875e3f1}{Supplementary Material} for the detailed demographics)}. The qualitative analysis is based on the 30 participants \revision{(16 male, 14 female; mean age = 20.878, SD = 1.899)} who were selected for and completed the post-study interviews \revision{(full demographic information available in the \href{https://osf.io/k3gvh/overview?view_only=0a2b40f80e5d4ec9b3fb96838875e3f1}{Supplementary Material})}. To aid interpretation, Table \ref{tab:thematic_analysis_summary} summarizes the three overarching themes identified in our analysis, their core insights, and the research questions they address.

\begin{table*}[ht]
\small
    \centering
    \begin{tabularx}{.8\textwidth}{XXX}
        \toprule
         \textbf{Theme} & \textbf{Related RQs} & \textbf{Theme insights}  \\
         \midrule
         Perceptual Assessment of the Virtual Peer & \textbf{RQ3} (impact of appearance) and \textbf{RQ4} (assessment of appearance, behavior, believability, and perceived deception). & Participants relied on visual appearance to assess the virtual peer and position it within their social landscape. The more human-like agent (SEPH) was often perceived as more similar or familiar, sometimes supporting a sense of identification, whereas more cyborg-like design (SEPC) did not evoke such reactions. Appearance also served as a key cue for categorizing partners -- helping participants immediately distinguish AI peers from human peers and shaping the expectations they brought into the interaction. \\
         \midrule
         Believability and Deception & \textbf{RQ4} (users’ evaluation of behavior, communication, believability, and perceived deception). & Believability depended on coherence and role-fit rather than human-likeness. Participants evaluated Alex based on whether its tone, responsiveness, and behavioral cues matched what they expected from an AI companion -- accepting signs of artificiality but reacting negatively when messages felt unrealistic, pressured, or poorly timed. Deception was not a concern; instead, participants judged the agent as a believable AI when its behavior aligned with its role. \\
         \midrule
         Relational Dynamics with Peers & \textbf{RQ1} (SEP ability to provide relatedness support) and \textbf{RQ2} (comparison between AI and human peers). & This theme captures both how social relationships were formed and how social comparison dynamics unfolded. Relationships unfolded differently across conditions: human peers offered the potential for authentic connection but also introduced relational risks (no replies, mismatch of effort), while AI peers provided stable, low-stakes, but bounded forms of companionship and support. Social comparison dynamics varied widely, with competence experienced socially (HUM) or reflectively (SEP). \\
         \bottomrule
    \end{tabularx}
    \caption{Summary of the three overarching themes identified through thematic analysis, including their analytic insights and their alignment with the study’s research questions.}
    \Description{This table synthesizes how each theme captures a distinct aspect of participants’ experiences with human and AI peers, while also clarifying how the qualitative findings contribute to answering RQ1-RQ4.}
    \label{tab:thematic_analysis_summary}
\end{table*}

\subsection{RQ1: To what extent can AI agents effectively provide relatedness support to promote and sustain physical activity?}
\label{sub:results_rq1}
\begin{table*}[ht]
    \centering
    \begin{tabular}[t]{llrrrll}
        \toprule
        \textbf{DV} & \textbf{Predictor} & \textbf{$\beta$} & \textbf{SE} & \textbf{t} & \textbf{p} & \textbf{95\% CI}\\
        \midrule
        \multirow{3}{*}{Phase} & Intercept & 0.00 & 0.07 & 0.01 & = 0.989 & {}[-0.14, 0.14]\\
        & Deployment & -0.11 & 0.02 & -5.56 & < .001 & {}[-0.15, -0.07]\\
        & Follow-up & 0.22 & 0.02 & 10.14 & < .001 & {}[0.18, 0.26]\\
        \addlinespace
        \multirow{3}{*}{Condition} & HUM & 0.10 & 0.10 & 1.02 & = 0.310 & {}[-0.10, 0.30]\\
        & SEPC & -0.12 & 0.10 & -1.15 & = 0.250 & {}[-0.32, 0.08]\\
        & SEPH & 0.07 & 0.11 & 0.67 & = 0.502 & {}[-0.14, 0.28]\\
        \addlinespace
        \multirow{6}{*}{Phase x Condition} & Deployment x HUM & -0.04 & 0.03 & -1.50 & = 0.135 & {}[-0.10, 0.01]\\
        & Follow-up x HUM & -0.05 & 0.03 & -1.51 & = 0.132 & {}[-0.10, 0.01]\\
        & Deployment x SEPC & -0.10 & 0.03 & -3.62 & < .001 & {}[-0.16, -0.05]\\
        & Follow-up x SEPC & -0.12 & 0.03 & -4.11 & < .001 & {}[-0.18, -0.07]\\
        & Deployment x SEPH & -0.10 & 0.03 & -3.31 & < .001 & {}[-0.15, -0.04]\\
        & Follow-up x SEPH & -0.20 & 0.03 & -6.15 & < .001 & {}[-0.26, -0.14]\\
        \bottomrule
    \end{tabular}
    \caption{Linear mixed-effects model predicting weekly steps. Reported are unstandardized coefficients ($\beta$), standard errors (SE), t-values, p-values, and 95\% CIs.}
    \label{tab:lmer_results}
\end{table*}
Results from the linear mixed-effects model predicting weekly step counts are summarized in Table \ref{tab:lmer_results}. An Analysis of Deviance (Type II Wald $\chi^2$ test) on our LMER model for weekly steps revealed a significant main effect of Phase ($\chi^2$(2) = 811.78, p < .001) and a significant Phase × Condition interaction ($\chi^2$(6) = 49.10, p < .001). There was no significant main effect of Condition ($\chi^2$(3) = 7.26, p = .064).
\begin{figure*}[ht]
    \centering
    \includegraphics[width=.8\textwidth]{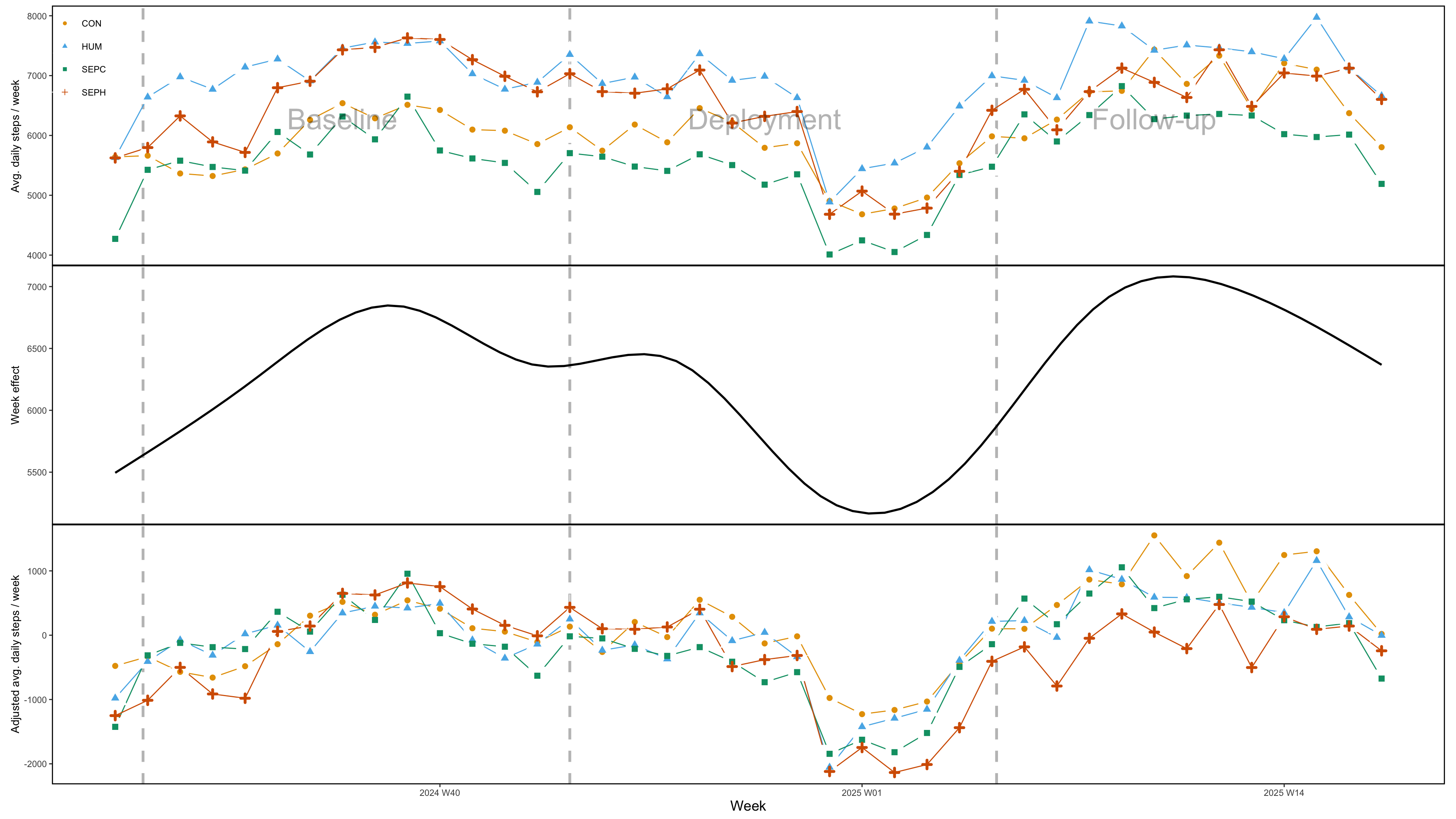}
    \caption{Weekly average step counts across conditions (top panel), with overall week effect (middle panel) and condition-adjusted residuals (bottom panel). Vertical dashed lines indicate the three study phases: Baseline, Deployment, and Follow-up. Human-Human pairs (HUM) consistently maintained higher step counts than cyborg peers (SEPC) during Deployment and Follow-up, while both SEP conditions showed trajectories closer to Control.}
    \Description{Evolution of step counts across the study. The top panel shows weekly average daily steps for each condition (CON, HUM, SEPC, SEPH) during baseline, deployment, and follow-up phases. The middle panel shows the overall week effect as a smooth curve, reflecting temporal fluctuations independent of condition. The bottom panel shows adjusted weekly average steps (controlling for week effects), highlighting differences between conditions over time.}
    \label{fig:steps_analysis}
\end{figure*}
The significant Phase × Condition interaction indicates that the trajectory of participants’ weekly steps over time differed by condition. To probe this interaction, we conducted post-hoc pairwise comparisons of estimated marginal means with a Tukey adjustment. During the Deployment phase, participants in the HUM condition recorded significantly more weekly steps than those in the SEPC condition (est. = 0.28, SE = 0.10, p = .032). A similar pattern emerged at Follow-up, where HUM participants again outperformed SEPC (est. = 0.30, SE = 0.10, p = .019). No other pairwise contrasts were statistically significant in any phase. Figure \ref{fig:steps_analysis} illustrates these trajectories, showing that Human-Human pairs maintained higher weekly step counts than cyborg peers during both Deployment and Follow-up.

\par Our thematic analysis helps to explain these trajectories through the overarching theme of Relational Dynamics with Peers.

\par \revision{This theme points to how peers became emotionally meaningful figures of comparison -- sometimes motivating, sometimes discouraging, and sometimes irrelevant -- revealing that comparison is not inherently energizing but deeply shaped by relational context. This dynamic was} particularly salient in the HUM condition, but participants’ reactions varied widely. For some, seeing their partner’s progress was motivating, even if they could not always keep up: (P24, HUM) \textit{``It pushed me to try to catch up, but unfortunately I had constraints in my life that didn’t always allow me to''}. Others described comparison more as curiosity than competition: (P26, HUM) \textit{``When the partners arrived, I compared a bit, just out of curiosity, to see how the other was evolving''}. Still, some participants reported little to no effect, such as one who admitted, (P30, HUM) \textit{``No, it really didn’t do much for me... it just didn’t matter''}.

\par These accounts suggest that while human peers often provided a strong source of comparison, their impact depended heavily on individual attitudes and circumstances, ranging from competitive drive to indifference.

\par \revision{In the SEP conditions, participants described comparison with an AI peer as a gentler, less emotionally charged practice -- useful for benchmarking activity but difficult to experience as real competition. Many SEP participants appreciated the light sense of challenge created by Alex’s step counts:} (P11, SEPH) \textit{``I liked the idea of competition between the chatbots and us, because it still gave me a sense of motivation to stay engaged''}. Others used the agent’s progress as a benchmark, not to compete directly, but to set daily goals: (P17, SEPC) \textit{``There was this kind of bar to know the number of steps I was doing each day... if I saw I was below, I told myself to go out and walk a bit more''}. Yet participants also recognized the limits of such comparisons. One reflected, (P3, SEPH) \textit{``Even if they told me every day how many steps the AI did, even knowing it was fake, maybe I’d feel in competition... but deep down there’s nothing behind it''}. The same participant described struggling with the very idea of exertion for an AI peer (P3, SEPH) \textit{``It felt strange, because since it’s virtual it doesn’t literally walk or run... I had trouble convincing myself I was competing with it''}. Another participant made this skepticism explicit: (P23, SEPC) \textit{``I don’t consider that the bot can do physical activity; that’s why I said those are arbitrary numbers, there isn’t a person walking in front of me''}.

\par In sum, the qualitative findings for RQ1 suggest that both human and AI peers shaped physical activity through social comparison, but in different and sometimes ambivalent ways. Human peers often motivated through authentic effort and accountability, yet participants’ reactions ranged from competitive drive to indifference depending on individual circumstances. SEP peers, by contrast, provided reliable encouragement and goal-setting cues, and while some participants appreciated the idea of simulated competition, others struggled to attribute genuine exertion to an AI partner. This variability underscores that humans and agents influence behavior through distinct but imperfect mechanisms -- competition and accountability for HUM, consistency and encouragement for SEP.

\par These dynamics highlight that the more profound differences between human and AI peers may lie in how they foster social connection and support, rather than in activity outcomes alone. We therefore turn to the qualitative and quantitative results on perceived competence, relatedness, social presence, and working alliance.

\subsection{RQ2: How do AI agents compare to humans in providing relatedness support to promote and sustain physical activity? and RQ3: How does the visual appearance of an AI agent, specifically a human figure versus a cyborg, affect its ability to provide relatedness support?}
\label{sub:results_rq2_rq3}
\begin{table*}[ht]
    \centering
    \begin{tabularx}{.75\textwidth}{llXXX}
        \toprule
        \textbf{DV} & \textbf{Effect} & \textbf{F(df1, df2)} & \textbf{p} & \textbf{$\eta^2$\_G}\\
        \midrule
        \multirow{4}{*}{IMI Perceived Competence} & (Intercept) & F(1, 125) = 4835.32 & < .001 & 0.967\\
        & Condition & F(3, 125) = 1.54 & = 0.207 & 0.027\\
        & Time & F(2, 250) = 5.85 & = 0.003 & 0.012\\
        & Time x Condition & F(6, 250) = 1.89 & = 0.083 & 0.011\\
        \addlinespace
        \multirow{4}{*}{IMI Relatedness} & (Intercept) & F(1, 125) = 10023.98 & < .001 & 0.976\\
        & Condition & F(3, 125) = 2.20 & = 0.091 & 0.026\\
        & Time & F(2, 250) = 2.70 & = 0.069 & 0.011\\
        & Time x Condition & F(6, 250) = 1.08 & = 0.373 & 0.013\\
        \addlinespace
        \multirow{4}{*}{Social Presence} & (Intercept) & F(1, 91) = 1156.42 & < .001 & 0.897\\
        & Condition & F(2, 91) = 4.22 & = 0.018 & 0.060\\
        & Time & F(2, 182) = 0.50 & = 0.610 & 0.002\\
        & Time x Condition & F(4, 182) = 1.76 & = 0.138 & 0.012\\
        \addlinespace
        \multirow{4}{*}{WAI Bond} & (Intercept) & F(1, 92) = 914.61 & < .001 & 0.882\\
        & Condition & F(2, 92) = 21.98 & < .001 & 0.265\\
        & Time & F(2, 184) = 4.44 & = 0.013 & 0.012\\
        & Time x Condition & F(4, 184) = 0.57 & = 0.683 & 0.003\\
        \bottomrule
    \end{tabularx}
    \caption{Mixed ANOVA results for each questionnaire's dependent variable, including main effects of Condition, Time, and the Time × Condition interaction. Reported are F statistics, p values, and generalized eta squared ($\eta^2$G). Post-hoc pairwise comparisons are reported in the \href{https://osf.io/dwjzk/?view_only=1b3a38027eed4d5290f321a0552dec08}{Supplementary Material}.}
    \label{tab:manova_questionnaires}
\end{table*}

\begin{figure*}[ht]
    \centering
    \includegraphics[width=.7\textwidth]{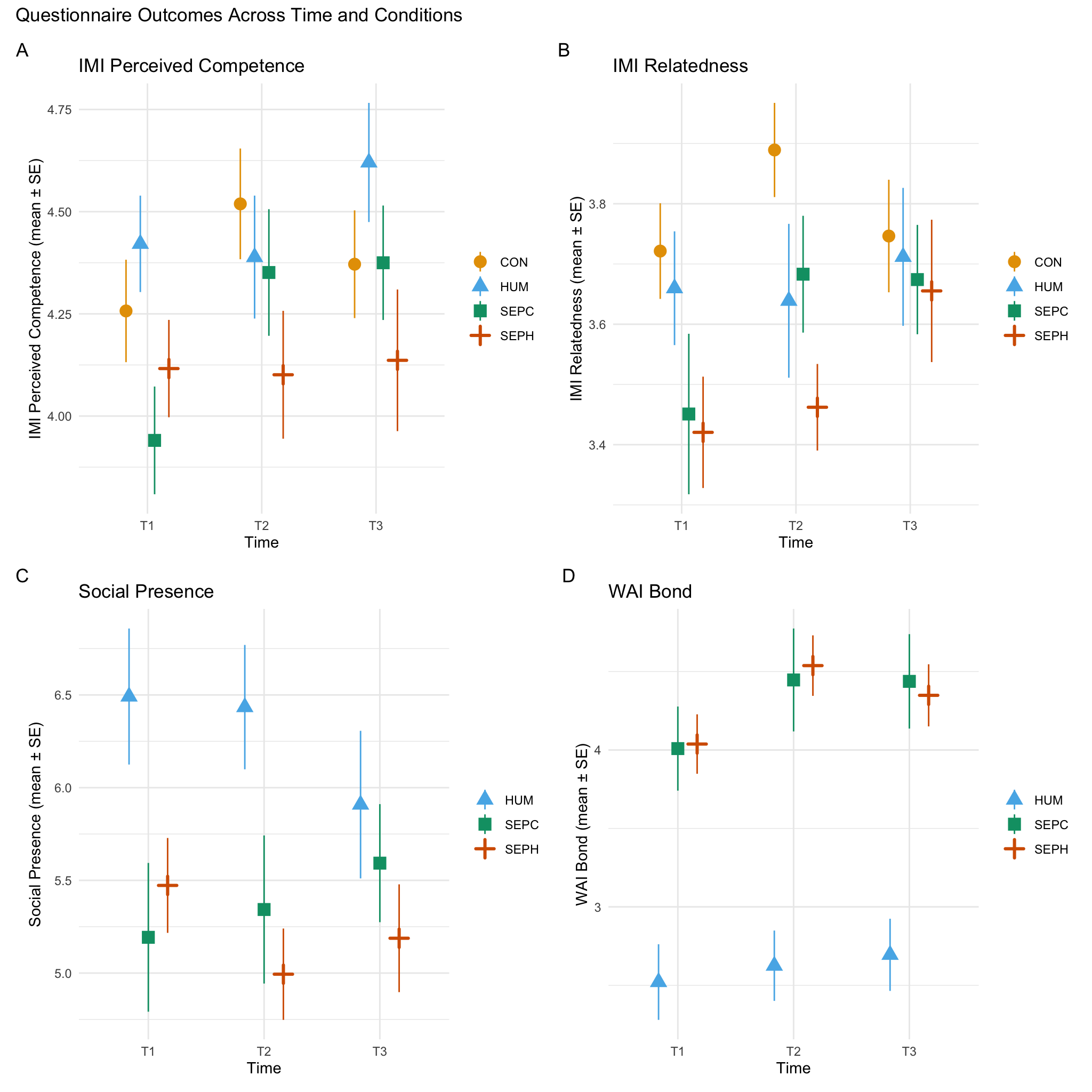}
    \caption{Questionnaire outcomes across time and conditions. (A) IMI Perceived Competence, (B) IMI Relatedness, (C) Social Presence, (D) WAI Bond. Error bars represent ±1 SE.}
    \Description{Questionnaire outcomes across time and conditions. Four panels display mean scores (±SE) at three time points (T1, T2, T3). (A) Perceived competence shows relatively stable scores across conditions, with HUM trending slightly higher. (B) Relatedness is generally higher in the control condition, while AI peers (SEPC, SEPH) score lower. (C) Social presence is highest in the HUM condition and consistently lower for AI peers. (D) Working Alliance Bond (WAI Bond) is markedly higher for AI peers (SEPC, SEPH) compared to human peers.}
    \label{fig:questionnaires_mean_se}
\end{figure*}

The results of the Mixed ANOVAs for perceived competence, relatedness, social presence, and working alliance are presented in Table \ref{tab:manova_questionnaires}. Across measures, we observed significant effects of both time and condition, with the strongest condition effects emerging for social presence and working alliance. Post-hoc Tukey-adjusted comparisons indicated that Human-Human pairs reported greater social presence than AI peers, whereas both SEP conditions scored higher on working alliance than HUM participants. These results are illustrated in Figure \ref{fig:questionnaires_mean_se}, which shows that while perceived competence increased steadily across phases in all conditions, relational outcomes diverged: human peers primarily fostered social presence, while AI peers established stronger working alliances. Together, these findings suggest that humans and AI support motivation through complementary relational pathways.

\subsubsection{Perceived Competence}
\label{subs:results_rq2_rq3_pc}
The analysis for Perceived Competence revealed a significant main effect of Time (F(2, 250) = 5.85, p = .003\revision{, $\eta^2$ = .012}). As illustrated in Figure \ref{fig:questionnaires_mean_se}A, perceived competence generally improved over time, though patterns varied by condition. SEPC participants showed the largest increase from baseline to follow-up, while HUM participants ended with the highest competence ratings at Time 3. In contrast, SEPH remained stable across all phases.

\par \revision{The theme of Relational Dynamics with Peers further helps clarify competence-related experiences: participants in the SEP conditions described how comparison with Alex became a more reflective and personally anchored process -- leading them to question the meaning of ``beating'' an AI while still benefiting from its steady, empathic support. Some participants found that the 10,000-step benchmark or ``winning'' against the agent felt arbitrary or not representing a meaningful goal; as one noted, (P1, SEPH) \textit{``For me, it’s not a sufficient goal [either reaching 10,000 steps or beating Alex]. I need to reach a place that is beautiful for me; the goal has to be something beyond 10,000 steps [or Alex]... I am already commuting one and a half to two hours to the university, fitting in 10,000 steps is hard''}}. Others reflected more positively on the agent’s support in achieving realistic objectives. One participant described moments of disappointment when they fell short, but also appreciated the agent’s empathy: (P8, SEPH) \textit{``I was quite disappointed in myself... but I think I stayed realistic about not being able to achieve all my goals every week. That’s also where the agent was empathetic, when I didn’t manage, it was still just a small disappointment''}. The same participant highlighted how the agent helped sustain motivation over time: (P8, SEPH) \textit{``Overall it was an increase, even if there were variations... she helped me stick to my goals, especially after our conversations''}.

\par \revision{Within the same theme, participants in the HUM condition framed competence as a socially grounded experience -- derived from outperforming a real peer but easily undermined when that peer offered too little or too much challenge. Some described genuine satisfaction when they exceeded their partner’s activity levels:} (P20, HUM) \textit{``During periods when I was more active, from time to time I beat them... it was a small personal satisfaction, but nothing extraordinary; it wasn’t my weekly goal''}. Yet others were frustrated by a lack of challenge when their peer was consistently weaker: (P27, HUM) \textit{``I really had the impression there was nothing in front of me. At least if there had been more, it would have motivated me... out of curiosity it would have pushed me to go on the app more''}.

\par These accounts suggest that perceived competence was reinforced through different pathways: for HUM participants, through relative performance against a peer -- though this could motivate or frustrate depending on parity -- and for SEP participants, through sustained encouragement and empathy that normalized setbacks while supporting gradual progress toward goals.

\subsubsection{Perceived Relatedness, Working Alliance, and Social Presence}
\label{subs:results_rq2_rq3_pr_wai_sp}
A Mixed ANOVA on the IMI Perceived Relatedness scores showed \revision{a non-significant main effect of Condition}, \revision{F(3, 125) = 2.20, p = .091, $\eta^2$ = .026}. As shown in Figure \ref{fig:questionnaires_mean_se}B, perceived relatedness remained relatively stable across all conditions. SEPC and SEPH showed modest increases from baseline, with SEPH participants reporting slightly higher scores by the final phase. In contrast, HUM participants’ ratings remained flat throughout.

\par For the WAI Bond scores (Figure \ref{fig:questionnaires_mean_se}D), participants in the HUM condition consistently reported lower working alliance bond compared to both SEP conditions. HUM scores remained around 2.5-2.7 across phases, whereas SEPC and SEPH participants reported values near 4.0-4.5. Post-hoc tests confirmed that HUM was significantly lower than both SEPC and SEPH (p < .001), while the two SEP-driven conditions did not differ significantly from one another.

\par On the Social Presence Scale (Figure \ref{fig:questionnaires_mean_se}C), participants in the HUM condition consistently reported higher social presence than those paired with SEP agents. Although the omnibus test indicated only a marginal main effect of Condition (p = .059), the descriptive pattern was robust, with HUM scores remaining about one point higher than SEPC and SEPH across all phases.

\par \revision{Within the main theme of Relational Dynamics with Peers, participants described two distinct relational trajectories: human peers offered the possibility of genuine connection but also the risk of relational disappointment, whereas the AI peer provided steady friendliness that felt supportive yet fundamentally limited in depth. In the HUM condition,} some participants described polite and respectful exchanges that created a minimal sense of connection: (P12, HUM) \textit{``He was nice, so I sent him messages more often... but in the end there weren’t many expectations, just respectful exchanges in the context of the experiment''}. Others reported little engagement, either because no one initiated conversation (P26, HUM; \textit{``I didn’t make the first move either, so we never talked''}) or because they had no interest in forming a relationship (P27, HUM; \textit{``Honestly I wasn’t looking for much, even nothing, in terms of discussion''}). Attempts at interaction could also lead to frustration, as one explained: (P4, HUM) \textit{``You feel stupid when you try too hard and just get curt replies... it’s like reaching out and getting rejected''}. By contrast, SEP participants described a more consistent, if limited, sense of connection. For some, the agent felt supportive but not like a ``real'' partner: (P3, SEPH) \textit{``It was a bit like a relationship where she supported me to take more steps, but nothing more''}. Others acknowledged the friendly tone but could not feel genuinely closer: (P7, SEPH) \textit{``It wasn’t a friend or a close person... despite all the kind words I didn’t get more motivated''}. Some even imagined greater closeness in different contexts, as one reflected: (P11, SEPH) \textit{``If I had had this as a kid, I would have talked to the chatbot like a real human, almost like a buddy''}. In a few cases, participants in SEPC reported a sense of growing closeness over time, even seeking advice from the agent.

\par Social presence followed a similar pattern. When human peers were responsive, participants described a stronger feeling of ``someone there,'' but this could collapse when partners were disengaged, leaving some to describe a lack of presence or even rejection (P4, HUM). With SEP peers, presence was steadier but often described as artificial. One participant emphasized that, despite the agent’s conversational cues, (P7, SEPH) \textit{``it was just an AI trying to motivate me''}, while another explained that self-identification with the SEPH avatar made the peer feel more immediate and present.

\par Perceptions of bond were shaped less by authenticity than by consistency. HUM participants often did not expect or seek a deeper bond (P27, HUM; \textit{``I wasn’t expecting to become friends with my partner''}). In contrast, SEP participants sometimes described a supportive alliance, especially when the agent responded empathetically to struggles: (P8, SEPH) \textit{``I was disappointed in myself... but that’s where the agent was empathetic, it understood even if I failed''}. At the same time, limits such as repetitive messages or poorly timed reminders reduced the sense of trust.

\par \revision{Within the main theme of Perceptual Assessment of the Virtual Peer, participants’ accounts showed that visual appearance shaped how ``peer-like'' or relatable the agent felt -- those interacting with the human-like avatar were more likely to describe it as familiar, which helped them engage with it as a low-stakes companion.} These participants in the SEPH condition also reported self-identifying with their peer, explaining that it looked like them or shared physical characteristics -- a phenomenon not observed with the cyborg agent. A desire for personalization was echoed by some in the SEPC condition, who mentioned wanting more features to shape the agent to their liking. This consistent, low-stakes interaction helps explain why the SEPH condition fostered a greater sense of relatedness than being alone, while the agent's obvious limitations explain why it could not match the authentic social presence and bond of a successful human dyad.

\par Overall, the results point to complementary strengths: humans fostered authentic presence and depth when engagement occurred, while AI peers offered consistent encouragement and a reliable, if limited, sense of bond (Figure \ref{fig:questionnaires_mean_se}).

\subsection{RQ4: How do users actively assess and evaluate the virtual peers' visual appearance, behavior, and communication style, considering both their level of believability and the potential for perceived deception within the context of a physical activity intervention?}
\label{sub:results_rq4_deception}
\revision{Within the main theme of Believability and Deception, participants’ accounts revealed that believability depended less on whether the agent appeared human and more on whether its behavior coherently matched what they expected from an AI companion. Visual appearance served as an initial cue, shaping early impressions but rarely dominating long-term evaluation. \revision{The theme of Perceptual Assessment of the Virtual Peer further helps understand this point.} Some participants appreciated the presence of an avatar, which added familiarity or warmth: (P11, SEPH) \textit{``There was my avatar and then the chatbot’s avatar, and that added something extra, I found it nice''} Others interpreted the simplified or gender-neutral design as an intentional way to make the agent easier to engage with: (P7, SEPH) \textit{``I think it was maybe a way to make it easier to form a link... removing masculine or feminine traits so it’s more neutral.''} At times, participants projected gender or personality traits onto the agent based on hairstyle or expressive style (P2, SEPC).}

\par \revision{Still, as it appears in the theme of Believability and Deception, many explained that appearance quickly faded into the background as the interaction unfolded, while conversational style ultimately became the core determinant of believability. Participants highlighted recurring signals of artificiality, such as repetitive structure (P23, SEPC) or exaggerated expressive punctuation (P19, SEPH). Others noted shallow conversational depth or looping responses (P9, SEPC). Yet the agent also produced moments of warmth or empathy that felt surprisingly human-like: (P25, SEPC) \textit{``It was nice to talk to someone who respected and motivated you,''} or (P8, SEPH) \textit{``When I said I didn’t have time, it understood and responded like a human might have.''} This blend of robotic consistency and occasional emotional resonance meant that participants rarely confused Alex for a human, but readily accepted it as a believable AI partner.}

\par \revision{Expectations played a significant role in shaping these judgments. Some participants preferred a more neutral or restrained tone (P10, SEPC), while others hoped simply for supportive communication without extremes (P13, SEPH). Believability faltered when messages violated expectations of realism or appropriateness -- for instance, when the agent insisted on walking despite genuine constraints (P22, SEPC) or when reminders felt poorly timed or overly insistent during stressful periods (P1, SEPH).}

\par \revision{Together, these reflections show that believability less about deception or mistaken identity, but about coherence and role-fit: participants evaluated the agent by whether its style, tone, and responsiveness aligned with what they believed an AI companion should be. When that coherence was maintained, the agent was treated as a reliable, if bounded, partner; when it slipped, participants expressed frustration or withdrew from interaction.}

\section{Discussion}
\label{sec:discussion}

%
%
%
%
%

Our six-month longitudinal study investigated the capacity of AI-driven Simulated Exercising Peers (SEPs) to provide social support for physical activity, directly comparing their impact against human peers and a control condition. To synthesize the quantitative outcomes, Table \ref{tab:hypotheses_table} summarizes the hypothesis testing results across behavioral and relational measures. Taken together, these findings reveal a fascinating paradox in human-agent interaction, challenging a simple view of an ``authenticity gap.'' While human partners were superior at creating a feeling of social presence, AI partners were more effective at establishing a strong working alliance. This suggests that human and AI partners excel in different relational dimensions: humans provide a richer sense of authentic social connection, while AI provides a more functional, reliable, and task-focused collaboration.

\begin{table*}[ht]
    \centering
    \begin{tabularx}{\textwidth}{cXXXc}
         \toprule
         \textbf{Hypothesis} & \textbf{Description} & \textbf{Omnibus test result} & \textbf{Post-hoc/Contrast result} & \textbf{Support} \\
         \midrule
         \textbf{H1a} & Increased Steps (from Pre to Dep): Human-Human/Human-SEP $>$ Control & Interaction: $\chi^2(6)=49.10$, $p<.001$ & SEP vs. CON interactions: $p\leq.001$. HUM vs. CON interaction: $p=.135$ & \textbf{Partially supported} \\ 
         \addlinespace
         \textbf{H1b} & Decreased Steps (from Dep to Post): Human-Human/Human-SEP $<$ Control & Interaction: $\chi^2(6)=49.10$, $p<.001$ & SEPH vs. CON interaction: $p=.001$. Others non-significant. & \textbf{Partially supported} \\
         \addlinespace
         \textbf{H1c} & Increased Steps (from Pre to Post): Human-Human/Human-SEP $>$ Control & $\chi^2(6)=49.10$, $p<.001$ & All EMM comparisons vs. Control were non-significant. & \textbf{Not supported} \\
         \addlinespace
         \textbf{H2a} & Relatedness: Human-SEP $>$ Human-Human & Condition: $F(3,125)=2.20$, $p=.091$ & Comparison was not significant & \textbf{Not supported} \\
         \addlinespace
         \textbf{H2b} & Competence: Human-SEP $>$ Human-Human & Condition: $F(3,125)=1.54$, $p=.207$ & N/A (Omnibus not significant) & \textbf{Not supported} \\
         \addlinespace
         \textbf{H2c} & Social Presence: Human-SEP $>$ Human-Human & Condition: $F(2,91)=4.22$, $p=.018$ & Opposite effect found: HUM $>$ SEPC ($p=.004$) \& HUM $>$ SEPH ($p<.001$). & \textbf{Not supported} \\
         \addlinespace
         \textbf{H2d} & WAI Bond: Human-SEP $>$ Human-Human & Condition: $F(2,92)=21.98$, $p<.001$ & SEPC and SEPH scores were significantly higher than HUM scores. & \textbf{Supported} \\
         \addlinespace
         \textbf{H3a} & Relatedness: SEPH $>$ SEPC & Condition: $F(3,125)=2.20$, $p=.091$ & SEPH vs. SEPC: $p=1.000$ & \textbf{Not supported} \\
         \bottomrule
    \end{tabularx}
    \caption{Summary of hypothesis testing results, showing which predictions were supported, partially supported, or not supported across behavioral and relational outcomes.}
    \label{tab:hypotheses_table}
\end{table*}

\subsection{The partnership paradox: social presence vs. working alliance}
\label{sub:discussion_authenticity_gap}
Our findings expose a paradox at the heart of technology-mediated support: while human peers excelled at evoking authentic social presence, AI peers proved more effective at sustaining a reliable working alliance. Human connections were often experienced as authentic but inconsistent. The richness of mutual effort and the sense of ``someone real'' created moments of strong presence, yet this was undermined when peers disengaged or when social friction emerged. These observations align with prior studies that highlight the role of peers' mutual accountability and shared experience in sustaining physical activity \cite{bouten_students_2025,bong_good_2023}. At the same time, they also echo findings that peer influence can be detrimental, leading to disengagement or frustration when differences in involvement or commitment become apparent \cite{barta_similar_2023}. In contrast, AI peers never conveyed authenticity in the same way, but they consistently provided encouragement, availability, and a non-judgmental stance, consistent with prior work showing that conversational agents are often valued for their dependable and supportive presence \cite{bickmore_establishing_2005,bickmore_its_2005,sillice_using_2018}. This reliability proved sufficient to establish a stronger working alliance, even if the bond felt less ``real,'' echoing earlier work on relational agents capabilities to foster working alliance bonds \cite{bickmore_establishing_2005}.

\par This paradox offers a refinement to SDT, which often frames relatedness as primarily fulfilled through close human connection \cite{deci_what_2000,ryan_intrinsic_2020}. Our results suggest that relatedness can also be scaffolded through reliability and consistency, even when authenticity is absent. Human peers fulfilled relatedness needs by offering genuine social presence, but this came with the unpredictability of real relationships. Similar dynamics have been noted in digital fitness contexts, where features that foster interactivity can increase users' sense of relatedness but at the same time reduce autonomy and competence, highlighting the double-edged nature of social connection in technology-mediated support \cite{molina_motivation_2023}. AI peers, by contrast, only partially satisfied relatedness, yet they sustained engagement by reliably reinforcing competence and providing steady encouragement, a pattern consistent with prior work that highlights agents' supporting capabilities and constant presence as central to their value \cite{bickmore_establishing_2005,bickmore_its_2005,sillice_using_2018}. In this way, our findings indicate that relatedness in mediated contexts is not binary -- authentic or artificial -- but rather multidimensional, encompassing both authenticity and reliability as distinct but complementary pathways to motivation. 

\par These differences also speak to ongoing debates around humanness and believability in human-agent interaction. Participants described the authenticity of human peers in terms of visible effort and reciprocity -- cues that AI could not convincingly reproduce. The AI peers' interactions were instead experienced as consistent but unmistakably artificial, marked by repetitive style or overly polished responses, echoing previous work showing that behavioral coherence rather than surface realism shapes social responses to agents \cite{rapp_human_2021}. This suggests that AI agents do not need to completely mimic humans to be believable, as their believability did not rest on appearance, but on the coherence of their behavior: while humans felt real because they offered genuinely human social presence, AI felt limited yet dependable in the support it provided. These dynamics illustrate how authenticity and reliability can each sustain engagement in different ways, pointing toward the divergent motivational pathways we examine next.


\par \revision{A further factor that may help explain why SEP participants found the AI peer especially relatable concerns relational tendencies intrinsic to LLM-driven agents. Instruction-tuned models are optimized through human feedback to provide consistently warm, agreeable, and supportive responses, often mirroring users' attitudes and maintaining a positive interpersonal tone -- a pattern documented as sycophancy or preference alignment in LLM behavior  \cite{sharma_towards_2024}. Research on empathic virtual agents similarly shows that simulated empathy and contingent emotional cues increase trust, social presence, and user engagement over time  \cite{paiva_empathy_2017}. In our study, these characteristics stood in sharp contrast to the interactions in the HUM condition (see Section \ref{sub:results_rq2_rq3}), where participants frequently reported delays, inconsistency, or even silent periods from their human partner. By comparison, Alex's responsiveness, emotional steadiness, and persistent availability created what participants described as a sense of ``reliable presence.'' Rather than reflecting a methodological artifact, this constant presence is a genuine design affordance of AI agent-based systems: their tone, consistency, and immediacy can be intentionally shaped through alignment and prompting, enabling AI peers to provide forms of relational support that are difficult to guarantee in human-human pairings, where availability and expressiveness naturally vary.}

\subsection{Divergent motivational pathways}
\label{sub:discussion_divergent}
While both human and AI peers supported participants in maintaining their exercise routines, they did so through markedly different motivational pathways. Human peers primarily motivated through social comparison and a sense of mutual responsibility. Knowing that another person was putting in effort created a benchmark, spurring competitiveness and a felt duty to respond in kind, an observation consistent with the core principles of Social Comparison Theory \cite{festinger_theory_1954}. These dynamics could be powerful drivers of engagement \cite{barta_similar_2023}, yet they also carried risks: unfavorable comparisons or perceived disengagement from a partner often led to frustration or discouragement as also observed in previous studies \cite{molina_motivation_2023}. Such volatility underscores the double-edged nature of social comparison in peer-based support.

\par AI peers did not function as credible comparison partners. Participants recognized that the AI was not exerting ``real effort,'' which limited its value for accountability or competition -- a challenge also noted in preliminary work on the difficulty of attributing effort to machines \cite{silacci_when_2025}. Instead, its motivational strength lay in reliability and encouragement. By offering steady feedback, normalizing setbacks, and being consistently available, AI peers created a supportive climate that buffered against social friction. This more stable but less intense pathway reflects the strengths of agents designed with a focus on relational features in fostering sustained, if less vivid, motivation \cite{lisetti_i_2013,bickmore_establishing_2005}.

\par Together, these contrasting pathways highlight the trade-offs inherent in human versus AI support. Human partners created peaks of motivation through authentic social comparison, but these peaks were interspersed with valleys of discouragement. AI peers offered flatter but steadier encouragement, producing a more sustainable motivational climate. Recognizing these divergent mechanisms is crucial for designing future systems that combine authenticity and reliability, harnessing the strengths of each without replicating their limitations.

\subsection{Role attribution and para-social boundaries}
\label{sub:discussion_role}
Beyond these mechanisms, participants also made sense of their partners by assigning them particular roles, which in turn shaped the depth and nature of the relationship. Human peers were often seen as workout buddies or comparable partners \cite{barta_similar_2023}, evoking a sense of shared effort and mutual accountability \cite{avraham_determinants_2024,bouten_students_2025}. Such role-based positioning could be motivating, but it also introduced vulnerability: when peers disengaged or failed to reciprocate, participants felt let down, underscoring the fragility of human roles in sustaining motivation.

\par AI peers, in contrast, were rarely mistaken for companions in the same sense. Participants readily acknowledged their artificial nature, which set clear boundaries around the relationship. Participants thus engaged with AI peers in a para-social way: they appreciated the AI's supportive stance but did not feel compelled to reciprocate care, since the relationship was understood to be fundamentally one-sided. This one-way dynamic also reduced the risk of disappointment. This resonates with recent research showing that para-social relationships with digital trainers are fostered when interactivity, authenticity, and companionship are perceived, and that authenticity is especially critical when the partner is an AI agent \cite{feng_virtual_2025}. A skepticism that also echoes recent evidence that, although people recognize AI companions as highly available and non-judgmental, they resist viewing such relationships as ``true'' because they lack the essential value of mutual caring \cite{oguz-uguralp_why_2025}. Instead, many participants cast AI peers in more constrained but dependable roles -- often likened to a coach, assistant, or supportive tool. This concurs with earlier work showing that users may prefer robots in coach-like positions rather than companion roles \cite{griffiths_exercise_2018}, and with participatory design studies of simulated exercising peers, where users explicitly distinguished between coach- and companion-like roles, highlighting how expectations of behavior and communication depended on the role an AI was assigned \cite{silacci_navigating_2025}. By recognizing the AI as non-human, participants adjusted their expectations accordingly, valuing consistency and encouragement without expecting genuine reciprocity. This emphasis on steady, non-judgmental support resonates with recent design work on LLM-based physical activity coaching, which advocates for facilitative rather than prescriptive guidance, contextual tailoring, and a supportive tone \cite{jorke_gptcoach_2025}.

\par Overall, these findings underscore that role attribution is central to how technology-mediated partners are experienced. Human partners offered authenticity but carried the risk of disappointment, whereas AI peers were dependable but limited in depth. Designing effective systems therefore requires aligning with these role expectations, balancing authenticity and reliability while setting clear relational boundaries.

\subsection{Design implications}
\label{sub:design_implications}
SEPs should lean into what AI does best: providing steady, non-judgmental, and always-available support. Our findings suggest that users trust AI when it is transparent about its limits and consistent in encouragement, but disengage when it tries to imitate human-like authenticity. Designing SEPs as dependable allies -- closer to coaches or assistants than companions -- can build sustainable trust. Making role expectations explicit helps reinforce these boundaries, ensuring that users see SEPs not as substitutes for human relationships but as reliable partners whose strength lies in consistency and availability.

\par At the same time, SEPs need to foster a sense of partnership by signaling effort and adapting motivational strategies to individual preferences. Although AI cannot truly share in physical exertion, design interventions can simulate co-experience by acknowledging progress, reflecting shared goals, and mirroring milestones. Participants also highlighted that motivation styles differ: some thrived on competition and comparison, while others preferred steady encouragement. SEPs that flexibly shift between competitive cues and supportive feedback can better sustain engagement. Looking forward, hybrid ecosystems that integrate the authenticity of human peers with the reliability of AI peers may offer the richest path, orchestrating complementary strengths to sustain long-term motivation.

\section{Limitations}
\label{sec:limitations}
While this study provides novel insights into the role of human and AI peers in supporting physical activity, several limitations should be acknowledged. Our sample consisted primarily of young adults in an academic setting, which may limit the generalizability of the findings to older adults, clinical populations, or individuals with different lifestyles and cultural backgrounds. Replicating these results in more diverse populations and across other health domains will be important to establish broader validity.

\par In terms of measurement, we focused on step counts, validated questionnaires, and interviews. These capture important behavioral and experiential dimensions but cannot fully reflect the complexity of physical activity motivation. Complementary approaches, such as ecological momentary assessments, physiological indicators, or richer digital trace data, could provide a more holistic picture of how motivation and engagement unfold in everyday life.

\par \revision{While our analyses referenced the 10,000-steps-per-day benchmark to contextualize daily activity, this figure has limited scientific grounding and originated as a popularized goal rather than an evidence-based threshold. Recent meta-analytic evidence indicates that substantial health and mortality benefits occur at lower activity levels -- approximately 8,000-10,000 steps for young adults \cite{paluch_daily_2022}. As our participants were young adults averaging around 7,000 steps per day (M = 7091, SD = 5773), their activity levels were below the range typically associated with optimal health benefits. Future studies may consider allowing users to set their own activity goals.}

\par Although our six-month intervention is substantially longer than most prior work, questions remain about sustainability over even longer periods. It is not yet clear whether the dynamics we observed (such as the partnership paradox) would persist, intensify, or diminish in year-long or indefinite deployments. Future research should therefore extend observation horizons to better understand the durability of human-AI peer effects.

\par Participant attrition also presents a limitation. Several individuals dropped out for reasons beyond our control, including exam periods, leaving without explanation, disengaging after encountering mobile application bugs, or discontinuing after receiving partial reimbursement at the end of the deployment phase. These patterns of attrition may have influenced both the sample size and the representativeness of the findings, underscoring the importance of developing retention strategies for long-term interventions in student populations.

\par Finally, the design of our system necessarily imposed constraints. The SEP peers were implemented as chat-based companions with simple avatars and limited adaptability. This provided control for experimental comparison, but also limited the richness of interaction. Exploring alternative embodiments, adaptive personalization, or multimodal feedback could yield different patterns of believability, presence, and social influence. As with all self-report methods, our questionnaires and interviews may also have been shaped by social desirability or recall bias; combining subjective accounts with behavioral or physiological indicators would strengthen confidence in future findings. \revision{Additionally, although, researcher monitoring allowed us to prevent inaccurate or unsafe feedback, Alex's responses were still subject to the well-known limitations of LLM-generated text. Future deployments should incorporate automated safety filters and domain-specific guardrails to further reduce the risk of inappropriate activity-related guidance.}

\par Despite these limitations, the study advances understanding of how human and AI peers support behavior change and provides a foundation for future work on adaptive, long-term, and ethically grounded designs.
\section{Conclusion}
\label{sec:conclusion}
This paper presented a six-month randomized controlled trial comparing human peers and AI-based simulated exercising peers in supporting physical activity. Our mixed-methods analysis revealed what we call the partnership paradox: human peers fostered stronger social presence, while AI peers established more reliable working alliances. These differences reflected divergent motivational mechanisms: humans inspired through authentic comparison and mutual responsibility, whereas AI peers motivated through steady encouragement and non-judgmental support.

\par These findings advance HCI research on human-agent interaction by showing that AI peers should not aim to replicate human authenticity, but rather complement it with reliability and coherence. By clarifying how believability, social presence, and motivational pathways intersect, this study highlights design opportunities for adaptive systems that flexibly balance competition and encouragement. Ultimately, the value of simulated exercising peers lies not in imitating human connection, but in serving as dependable, empathic collaborators that can help sustain physical activity and extend access to social support for healthier lifestyles at scale.

\begin{acks}
We sincerely thank all participants for their time, effort, and commitment throughout the study. We are grateful to the \anonymize{Labex organization from University of Lausanne} for their invaluable support in coordinating and scheduling participants, and we thank \anonymize{Marc-Olivier Boldi from the University of Lausanne} for his assistance with the statistical and power analyses. The development of our system was supported in part by the \anonymize{School of Management of Fribourg}, which provided funding for materials and equipment, and by the \anonymize{School of Engineering and Architecture of Fribourg,} which allowed us to use their App Store license. This research was funded by the \anonymize{University of Lausanne HEC Research Fund (Simulated Exercising Peers Towards Virtual Companionship to Support Physical Activity)}, which supported project development, the \anonymize{R\&D Fund of the School of Management Fribourg}, which covered participants’ expenses, and the \anonymize{Réseau de Compétence HES-SO (RCSO) Economie \& Management (Project SEP4PA)}, which supported the salary of one co-author.

\par ChatGPT (OpenAI) and Gemini (Google) were used under the authors’ supervision to improve grammar, flow, and writing polish; all ideas, analyses, and interpretations are solely those of the authors.
\end{acks}

\bibliographystyle{ACM-Reference-Format}
\bibliography{excero_bibliography}
\end{document}